\documentclass[10pt]{iopart}
\usepackage[latin1]{inputenc}
\usepackage[T1]{fontenc}
\usepackage{iopams}
\usepackage[dvipsnames,rgb,dvips]{xcolor}
\usepackage{graphicx}
\usepackage{dcolumn}
\usepackage{epstopdf}
\usepackage{siunitx}
\usepackage{mathrsfs}
\makeatletter
\def\amsbb{\use@mathgroup \M@U \symAMSb}
\makeatother
\usepackage{bbm}
\expandafter\let\csname equation*\endcsname\relax
\expandafter\let\csname endequation*\endcsname\relax

\usepackage{mathptmx}
\usepackage{amsmath,amssymb,amsfonts, MnSymbol}
\usepackage{mathrsfs}
\usepackage{bm}
\usepackage{hyperref}
\usepackage{bbold}
\usepackage{tikz}

\newcommand{\algn}[1]{\begin{align} #1 \end{align}}
\newcommand{\sbeqs}[1]{\begin{subequations} #1 \end{subequations}}

\newcommand{\css}[1]{\begin{cases} #1 \end{cases}}

\newcommand{\nn}{\nonumber}
\newcommand{\ee}{\ensuremath{\text{e}}}
\newcommand{\ed}{\ensuremath{\text{d}}}
\newcommand{\dd}[1]{\ensuremath{\tfrac{\text{d}}{\text{d} #1}}}

\newcommand{\Sint}{\ensuremath{S^\text{int}}}
\newcommand{\sint}{\ensuremath{\msc{S}^\text{int}}}

\newcommand{\tc}{\ensuremath{t_\text{c}}}
\newcommand{\tcenv}{\ensuremath{t^\msc{Q}_\text{c}}}
\newcommand{\bq}{\ensuremath{\beta_\text{q}}}
\newcommand{\bb}{\ensuremath{\bar\beta}}
\newcommand{\bc}{\ensuremath{\beta_\text{c}}}
\newcommand{\peq}{\ensuremath{P^\text{eq}}}
\newcommand{\peqq}{\ensuremath{P^\text{eq}_\text{q}}}
\newcommand{\veq}{\ensuremath{ \msc{V}^\text{eq} }}
\newcommand{\veqq}{\ensuremath{\msc{V}^\text{eq}_\text{q}}}
\newcommand{\msc}[1]{\ensuremath{\mathscr{#1}}}

\newcommand{\kb}{\ensuremath{k_\text{B}}}
\newcommand{\eqnlab}[1]{\label{eq:#1}}
\newcommand{\seclab}[1]{\label{sec:#1}}
\newcommand{\figlab}[1]{\label{fig:#1}}
\newcommand{\eqnref}[1]{\eqref{eq:#1}}
\newcommand{\Eqnref}[1]{Eq.~\eqref{eq:#1}}

\newcommand{\secref}[1]{\ref{sec:#1}}
\newcommand{\Secref}[1]{Sec.~\ref{sec:#1}}
\newcommand{\figref}[1]{\ref{fig:#1}}
\newcommand{\Figref}[1]{Figure~\ref{fig:#1}}
\newcommand{\Figsref}[1]{Figures~\ref{fig:#1}}

\begin{document}
\title{Landau theory for finite-time dynamical phase transitions}

\author{Jan Meibohm$^{1}$, Massimiliano Esposito$^2$}
\address{$^1$Department of Mathematics, King's College London, London WC2R 2LS, United Kingdom}
\address{$^2$Complex Systems and Statistical Mechanics, Department of Physics and Materials Science, University of Luxembourg, L-1511 Luxembourg, Luxembourg}

\begin{abstract}
We study the time evolution of thermodynamic observables that characterise the dissipative nature of thermal relaxation after an instantaneous temperature quench. Combining tools from stochastic thermodynamics and large-deviation theory, we develop a powerful theory for computing the large-deviation statistics of such observables. Our method naturally leads to a description in terms of a dynamical Landau theory, a versatile tool for the analysis of finite-time dynamical phase transitions. The topology of the associated Landau potential allows for an unambiguous identification of the dynamical order parameter and of the phase diagram. As an immediate application of our method, we show that the probability distribution of the heat exchanged between a mean-field spin model and the environment exhibits a singular point, a kink, caused by a finite-time dynamical phase transition. Using our Landau theory, we conduct a detailed study of the phase transition. Although the manifestation of the new transition is similar to that of a previously found finite-time transition in the magnetisation, the properties and the dynamical origins of the two turn out to be very different.
\end{abstract}
\section{Introduction}
Thermal relaxation is a fundamental process in statistical mechanics, with numerous applications in Nature and industry. Nonetheless, the kinetics of relaxation is well understood only close to equilibrium; within the quasistatic approximation~\cite{New01} and in the linear response regime~\cite{Ons31a,Ons31b,Kub57a,Kub57b}. As a major complication, far-from-equilibrium relaxation is characterised not only by the change of the instantaneous state of the system of interest, but also by its interaction with the surrounding environment. The dynamical interplay between system and environment manifests itself in macroscopically irreversible fluxes~\cite{Gro62} of thermodynamic quantities~\cite{Cri04,Imp07,Cor13,Cri17} that, in turn, determine the dissipative nature of the process. Consequently, far-from-equilibrium relaxation is a genuinely non-equilibrium problem that offers fascinating open questions, and a variety of unexpected phenomena.

A famous example of a relaxation anomaly is the Mpemba effect~\cite{Mpe69}, i.e., the faster cooling of an initially hotter system~\cite{Gre11,Ahn16,Lu17,Las17}. Other examples include asymmetries in the rates of heating and cooling~\cite{Lap20,Mei21b,Man21,Vu21}, as well as coarsening~\cite{Lif62,Cug15,Bra02}, ergodicity breaking~\cite{Bou92}, and ageing~\cite{Ber11} in glassy~\cite{Hun12} or phase-ordering~\cite{Lif62,Bra02} dynamics.

Often, but not always, anomalous relaxation phenomena are associated with the presence of equilibrium phase transitions, i.e., qualitative changes of the equilibrium state of a system under slow variation of the external parameters~\cite{Gol92,Cha95}. For example, in its original formulation~\cite{Mpe69}, the Mpemba effect corresponds to the (shorter) time it takes hot water to freeze, compared to cold water. Similarly, phase-ordering describes how a system condenses into its ordered phase starting in a disordered initial configuration~\cite{Lif62,Bra02}. The abrupt state changes associated with equilibrium phase transitions manifest themselves in singular points of thermodynamic quantities such as the free energy~\cite{Gol92,Cha95}.

The analysis of equilibrium phase transitions has led to the development of powerful methods, such as Landau theory~\cite{Lan37} or the renormalisation group~\cite{Kad66,Wil71a,Wil71b}, that have become standard tools of modern statistical mechanics. In particular at mean-field level, Landau theory provides a universal, widely model-independent picture of both continuous and first-order phase transitions in terms of the minima of a potential function, the so-called Landau potential. Although the importance of these methods in equilibrium statistical mechanics can hardly be overstated, their generalisation to non-equilibrium systems is not straightforward.

In the past decades, remarkable developments~\cite{Fre84,Gra84,Ton05,Ell07,Sei12,Cav14,Ber15,Pel21} in non-equilibrium statistical mechanics have led to conceptual generalisations of phase transitions to systems in non-equilibrium steady states~\cite{Der87,Vic95,Gar07,Her18,Shp18,Vro20,Mar20,Pro20,Ket21} and to dynamic observables~\cite{Meh08,Lac08,Ger11,Nya16,Jac10,Nya18,Nem19,Sun19,Laz19,Her20}, generating a pressing need for adequate theoretical tools to describe them. This demand has partly been met by a surge of new methods, based on, e.g., linear-response~\cite{Bai13,Fal16,Fre21}, optimal-control techniques~\cite{Che15b,Jac20} and machine learning~\cite{Whi20,Ros21,Yan22}, but also on non-equilibrium versions of Landau theory~\cite{Bae15,Smi18,Aro20,Hol22}.
 
In a recent Letter~\cite{Mei22a}, we reported another surprising relaxation phenomenon, a finite-time dynamical phase transition. This transition manifests itself in a finite-time cusp singularity~\cite{Ent02,Kul07,Ent10,Erm10,Red12,Fer13} of the probability distribution of the magnetisation of a mean-field magnet after an instantaneous quench of the temperature. In contrast to conventional phase transitions, this finite-time transition is induced by a change of the typical dynamics under variation of the observation \textit{time}. In other words, time takes the role of a control parameter, analogous to, e.g., the pressure in an equilibrium phase transition. Although unfamiliar in classical systems, similar finite-time transitions exist in conservative quantum systems as singularities of the Loschmidt echo~\cite{Hey13,Hey18}. The analysis in Ref.~\cite{Mei22a} is based on the standard Hamiltonian method~\cite{Fre84,Dyk94,Fen06} which allows one to compute the finite-time statistics of the state of the system in the thermodynamic limit.

In the present paper, we study the stochastic thermodynamics~\cite{Sei12,Bro15,Pel21} of the spin system in response to an instantaneous temperature quench from the ordered into the disordered phase of the magnet. We analyse the finite-time statistics of \textit{thermodynamic observables}, characterising the irreversibility of the relaxation process. Remarkably, the spins undergo another finite-time dynamical phase transition, associated with the heat exchanged between the spins and the bath, at a critical, finite time after the quench. As a consequence, the statistics of the exchanged heat develops a finite-time cusp that persists in the infinite-time limit. Despite the apparent similarity with the transition of the magnetisation~\cite{Mei22a}, we show that the new transition, associated with the exchanged heat, is of entirely different origin. As we shall describe in detail, the mechanisms that drive the two transitions, and their properties, are complementary. At the trajectory level, we show that the new transition is caused by a sudden switch of an optimal fluctuation of the spins with constrained initial and final points.

The analysis of finite-time transitions of thermodynamic observables requires a method that gives access to the time-dependent statistics of these observables in the thermodynamic limit. While existing non-equilibrium tools~\cite{Bae15,Che15b,Aro20,Jac20,Whi20,Ros21,Yan22} are typically tailored for steady states, finite-time approaches such as the mentioned Hamiltonian method~\cite{Fre84,Gra84,Dyk94,Fen06} are limited to phase transitions associated with the instantaneous state of the system~\cite{Mei22a}. Computing the finite-time statistics of thermodynamic observables thus requires a significant extension of the Hamiltonian method. In the course of this work, we develop such an extension by combining methods from stochastic thermodynamics~\cite{Sei12,Bro15,Pel21} and large-deviation theory~\cite{Fre84,Ell07,Hol08,Tou09}. In particular, we show that the finite-time statistics of thermodynamic observables naturally lend themselves to a description in terms of a dynamical Landau theory. The corresponding Landau potential is most useful in the presence of a dynamical phase transition, because its topology unambiguously identifies the dynamical phase diagram and its minima determine the dynamical order parameter.

More generally, our analysis reveals that finite-time dynamical phase transitions occur in a variety of ways for different observables within the same system. This hints towards the existence of such transitions in a much wider range of setups and it indicates that finite-time dynamical phase transitions of different nature are an integral part of the non-equilibrium statistical mechanics of relaxation. The extended Hamiltonian method including the dynamical Landau theory we develop here is a powerful tool for the identification and classification of finite-time dynamical phase transitions of systems in the thermodynamic or weak-noise limit.

The paper is organised as follows: In \Secref{background} we summarise the relevant background, including a brief review of the standard Hamiltonian method~\cite{Fre84,Gra84,Dyk94,Fen06} applied at finite time and of the main results of Ref.~\cite{Mei22a}. In \Secref{thermoobs} we give a detailed description of the extended Hamiltonian method, that allows us to compute the finite-time statistics of thermodynamic observables. This extension naturally leads to a dynamical version of Landau theory for thermodynamic observables at finite time. Finally, we use our theory to study the heat exchange of a mean-field magnet with its environment in \Secref{ftdpt}. We identify a finite-time dynamical phase transition associated with this observable and analyse and classify it in detail. In \Secref{conc} we draw our conclusions and describe future applications, as well as open questions.
\section{Background}\seclab{background}
In this section, we review the relevant background for our analysis. First, we define the Curie-Weiss model, a mean-field version of the Ising model, describe its equilibrium properties, and how we model its dynamics. We then explain the standard Hamiltonian method~\cite{Dyk94,Imp05,Fen06} and how it is used to analyse the statistics of the magnetisation of the model after a temperature quench. The post-quench dynamics of the magnetisation, previously discussed in Ref.~\cite{Mei22a}, exhibits a finite-time dynamical phase transition whose properties we review at the end of the section.
\subsection{Curie-Weiss model}
The Curie-Weiss model is a simplified caricature of a magnet where $N\to\infty$ Ising spins $\sigma_i=\pm1$ at sites $i=1,\ldots,N$ are coupled by an infinite-range, ferromagnetic interaction of strength $J/N>0$. The system is immersed in a heat bath at inverse temperature $\beta=1/(\kb T)$. Due to the mean-field nature of the interaction, we can write all microstates with equal numbers $N_{\pm}$ of spins in the states $\pm1$ in terms of the total magnetisation $M = N_+-N_-$. The constrained free energy $F(M)$ of the system at a given magnetisation $M$ reads~\cite{Cha95}
\algn{\eqnlab{epot}
 	F(M) = E(M) - \beta^{-1}\Sint(M)\,,
}
where $E$ the denotes internal energy
\algn{
	E(M) = -\frac{J}{2N}\left(M^2 -N\right)\,.
}
An additional coupling $-HM$ to an external field $H$ is omitted here. In this field-free version of the model, the internal energy $E(M)$ is entirely due to ferromagnetic interactions between the spins.

The dimensionless internal entropy $\Sint(M) = \ln\Omega(M)$ in \eqnref{epot} originates from the microscopic degeneracy of $M$:
\algn{
	\Omega(M)=\frac{N!}{[(N+M)/2]![(N-M)/2]!}\,.
}

State changes of the system are induced by thermal fluctuations of the heat bath, modelled by a stochastic dynamics for $M$. An arbitrary spin flip leads to a change $M\to M_\pm\equiv M\pm 2$ in the magnetisation. The probability $P(M,t)$ for finding the system in state $M$ at time $t$ obeys the evolution equation
\algn{\eqnlab{mastereqn}
	\dot P(M,t)=\sum_{\eta=\pm}\left[W_\eta(M_{-\eta})P(M_{-\eta},t)-W_\eta(M)P(M,t)\right]\,,
}
with Arrhenius-type rates $W_\pm(M)$ for the transitions $M\to M_\pm$, given by
\algn{\eqnlab{rates}
	W_\pm(M) =\frac{N\mp M		}{2\tau}\exp\left[ -\beta E_\pm(M)/2\right]\,.
}
Here, $\tau$ denotes the microscopic relaxation time for a single spin flip and $E_\pm(M) = E(M_\pm)-E(M)=-2J(\pm M+1)/N$ is the change of internal energy $E$ during the transition $M\to M_\pm$. The algebraic prefactor $(N\mp M)/2= N_{\mp}$ in \eqnref{rates} reflects that all $N_\mp$ microscopic transitions $\mp1\to\pm1$ are equivalent. Furthermore, the transition rates obey the spin-flip symmetry  $W_\pm(M)=W_\mp(-M)$ and the detailed-balance condition~\cite{Kam07}
\algn{\eqnlab{db}
 	\frac{W_\pm(M)}{W_\mp(M_\pm)}  = \frac{P^\text{eq}(M_\pm)}{P^\text{eq}(M)}= \exp\left[-\beta F_\pm(M)\right]\,,
}
where
\algn{\eqnlab{peq}
	P^\text{eq}(M) =  Z(\beta)^{-1}\exp\left[-\beta F(M)\right]\,,
}
denotes the equilibrium distribution with partition function $Z(\beta)$. In \eqnref{db}, the change in the constrained free energy $F_\pm(M)$ due to the transition $M\to M_\pm$ reads $F_\pm(M) = F(M_\pm) - F(M)$.
\subsection{Thermodynamic equilibrium}\seclab{equilibrium}
We now take the thermodynamic limit $N\to\infty$. To this end, we define the intensive magnetisation $m\equiv M/N$ per spin and the constrained free-energy density
\algn{\eqnlab{freeen}
	\msc{F}(\beta,m) \equiv \lim_{N\to\infty}F(Nm)/N = \msc{E}(m) - \beta^{-1} \sint(m),
}
with internal energy density
\algn{
	\msc{E}(m) = -Jm^2/2
}
and internal entropy per spin
\algn{\eqnlab{sint}
 	\sint(m)=-\sum_{\eta=\pm} \frac{1+ \eta m}2\ln\left(\frac{1+ \eta m}2\right)\,.
}
The equilibrium distribution in \eqnref{peq} written as a function of $m=M/N$, takes the large-deviation form~\cite{Ell07,Hol08,Tou09}
\algn{\eqnlab{ldp}
	P^\text{eq}(m) \smilefrown \exp\left[-N \veq(m)\right]\,,
}
with equilibrium rate function~\cite{Mei22a}
\algn{\eqnlab{veq}
	\veq(m)=\beta\left[\msc{F}(\beta, m)-\msc{\bar F}(\beta)\right]\,.
}
Here, the equilibrium free energy
\algn{\eqnlab{eqfreeen}
	\msc{\bar F}(\beta)=\lim_{N\to\infty}\frac1{\beta N}\ln Z(\beta)= \min_m \msc{F}(\beta, m)
}
arises as a consequence of the normalisation of $P^\text{eq}(M)$ in \eqnref{peq}, which ensures that $\veq(m)$ vanishes at its minima $\pm\bar m(\beta)$, i.e., $\veq[\bar m(\beta)]=0$. The magnetisation $\bar m(\beta)$ at the minima reflects the typical, most likely magnetisation that occurs at inverse temperature $\beta$ in the thermodynamic limit. According to \eqnref{ldp}, the probabilities for fluctuations of $m$ away from $\pm\bar m(\beta)$ are exponentially suppressed in $N$ at a rate given by $\veq(m)$. 

The shape of $\veq(m)$ reflects the macroscopic behaviour of the system at equilibrium. This macroscopic behaviour is driven by the tendency of the system to minimise its free energy, which results in a competition between energetic and entropic contributions in \Eqnref{freeen} as $\beta$ is varied. At small inverse temperatures, the entropy term $-\beta^{-1}\sint(m)$, minimised at $m=0$, dominates. In this case, $\veq(m)$ has a unique minimum at $\bar m(\beta)=0$, meaning that the contributions of up and down spins cancel each other and the system is said to be in a disordered state. Upon increasing $\beta$, however, the energy term $\msc{E}(m)$ in \Eqnref{freeen}, minimised at $m=\pm1$, becomes important and $\veq(m)$ changes its shape. In particular, at the critical inverse temperature $\bc=1/J$, $\veq(m)$ passes from single-well shape into that of a symmetric double-well, reflecting a continuous equilibrium phase transition~\cite{Lan37}. Consequently, for $\beta>\beta_\text{c}$ the magnetisation $\bar m(\beta)$ becomes finite and the system is said to be ordered, because $\bar m(\beta)>0$ implies that a dominant fraction of spins is aligned in either direction.

Figure~\figref{phase_diag}(a) shows the phase diagram of the Curie-Weiss model at vanishing external field, determined by the different topologies of $\veq(m)$ as $\beta$ is varied.
\begin{figure}
	\centering
	\includegraphics[width=9cm]{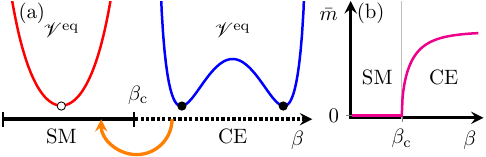}
	\caption{(a) Phase diagram for the Curie-Weiss model at vanishing external field, featuring the SM (solid black line) and CE (dotted line) phases, separated by the phase boundary at $\bc$. The equilibrium rate functions $\veq$ in the two phases are shown schematically in red and blue. The orange arrow indicates the direction of the disordering quench as described in the text. (b) Spontaneous magnetisation $\bar m(\beta)$ in the SM and CE phases.}\figlab{phase_diag}
\end{figure}
The phase diagram exhibits two phases: a disordered single-mode (SM) phase (solid black line) for $\beta<\bc$, where $\veq$ [red in \Figref{phase_diag}(a)] has a unique minimum (white bullet) and an ordered coexistence (CE) phase (dotted line) for $\beta>\bc$, where two degenerate, finite minima $\pm\bar m(\beta)$ (black bullets) of $\veq$ (blue) coexist. Close to $\bc$, the order parameter $\bar m(\beta)$, given by the minima of $\msc{F}$ and $\veq$, changes continuously from $\bar m(\beta) =0$ (disordered) to finite $\bar m(\beta)$ (ordered), as shown in \Figref{phase_diag}(b).

Objects like $\veq(m)$ or $\msc{F}(\beta, m)$, whose minima $\pm\bar m(\beta)$ specify the order (or disorder) of the system, are crucial tools for identifying equilibrium phase transitions, and are often summarised under the term ``Landau potentials''~\cite{Gol92,Cha95}. The corresponding Landau theory~\cite{Lan37}, aims to describe phase transitions by postulating phenomenological Landau potentials based on the microscopic symmetries of the problem. For the simple model we describe here, the functions $\veq(m)$ and $\msc{F}(\beta, m)$ can be derived explicitly, see \eqnref{veq} and \eqnref{freeen}.

The concept of Landau theory has proven useful also in non-equilibrium contexts~\cite{Bae15,Smi18,Aro20,Hol22,Mei22a}, where it serves to identify different non-equilibrium behaviours and dynamical order parameters. In \Secref{ftdpt}, we extend the notion of Landau theory to finite-time dynamical phase transitions of thermodynamic observables. Also in this case, the corresponding dynamical Landau potential, proves to be a powerful tool for the identification and the classification of the transition.
\subsection{Post-quench dynamics}\seclab{pqd}
At time $t<0$ we initialise the system in the CE phase at inverse temperature $\beta>\bc$, where $\veqq(m)$ has double-well shape and $\bar m(\beta)>0$. At $t=0$, we impose an instantaneous temperature quench $\beta\to\bq$ into the SM phase, i.e., $\bq<\bc$. Such a quench is said to be ``disordering'' as it forces the system to cross the phase boundary between the SM and CE phases [orange arrow in \Figref{phase_diag}(a)], inducing an order-to-disorder phase transition in the long-time limit~\cite{Mei22a}. Ergodicity ensures that $P(m,t)\to P^\text{eq}_q(m)\smilefrown\exp[-N\veqq(m)]$ as $t\to\infty$, where $\veqq(m)$ is the equilibrium rate function given in \eqnref{veq}, but at final inverse temperature $\bq$. For $t>0$, the time dependence of $P(m,t)$ is characterised by the large-deviation form
\algn{\eqnlab{ratem}
	P(m,t)\smilefrown\ee^{-NV(m,t)}\,,
}
with time-dependent rate function $V(m,t)$, whose evolution follows from an appropriate limit of \eqnref{mastereqn}. This limit is analysed using the standard Hamiltonian method~\cite{Fre84,Gra84,Dyk94,Fen06}.
\subsubsection{Standard Hamiltonian method}\seclab{ham}
When the transition rates $W_\pm$ are extensive in $N$, i.e.,
\algn{\eqnlab{ratelimit}
	w_\pm(q)=\lim_{N\to\infty}\frac{W_{\pm}(Nq)}{N}\,, \qquad 0<w_\pm(q)<\infty\,,
}
then \eqnref{mastereqn} transforms into a Hamilton-Jacobi equation for $V(m,t)$~\cite{Dyk94,Fen06},
\algn{\eqnlab{hj}
	0=\partial_t V(m,t) + \msc{H}[m,\partial_m V(m,t)]\,,
}
by substituting \eqnref{ratem} into \eqnref{mastereqn} and taking the large-$N$ limit. For the Curie-Weiss model, \eqnref{ratelimit} is satisfied and the dynamical Hamiltonian $\msc{H}$ reads~\cite{Imp05,Mei22a}
\algn{\eqnlab{hamiltonian}
	\msc{H}(q,p) &= w_+(q)\left(\ee^{2p}-1\right) + w_-(q)\left(\ee^{-2p}-1\right)\,,
}
including the $N$-scaled transition rates
\algn{\eqnlab{scaledrates}
	w_\pm(q) =\frac{1\mp q}{2\tau}\exp\left[\mp  \bq\msc{E}'(q) \right]=\frac{1\mp q}{2\tau}\exp\left(\pm \bq J q \right)\,.
}
The initial condition of \eqnref{hj} is given by the equilibrium rate function $\veq$ before the quench,
\algn{\eqnlab{vinit}
	V(m,0)=\veq(m)\,.
}

Solutions to the Hamilton-Jacobi equation~\eqnref{hj} are expressed in terms of the characteristics $q(s)$ and $p(s)=\partial_mV[q(s),s]$, $0\leq s\leq t$, that solve the Hamilton equations~\cite{Cou62}
\algn{\eqnlab{heom}
	\dot q(s) = \partial_p \msc{H}(q,p)\,,\quad \dot p(s) = -\partial_q \msc{H}(q,p)\,,
}
with the condition that $q(s)$ must end at $m$ at the given (observation) time $t$:
\sbeqs{\eqnlab{bound}
\algn{
	q(t) = m\,. \eqnlab{qtaubound}
}
The initial condition for $p(s)$ at $s=0$ follows from \eqnref{vinit}:
\algn{\eqnlab{p0bound}
	p(0) = \partial_mV[q(0),0] = \tfrac{\ed}{\ed m} \veq[q(0)]\,.
}
}
In order to solve the Hamilton-Jacobi equation~\eqnref{heom}, we evaluate the total derivative of $V[q(s),s]$ along $q(s)$:
\algn{\eqnlab{dVeqn}
	\dd{s}V[q(s),s] = \partial_mV[q(s),s]\dot q(s) + \partial_t V[q(s),s] = p(s)\dot q(s) - \msc{H}[q(s),p(s)]\,,
}
where we have used the definition of $p(s)$ and \eqnref{hj}. Integrating \eqnref{dVeqn} from $0$ to $t$ and using the boundary conditions~\eqnref{bound}, one obtains $V(m,t)$ as an integral over the characteristics $[q(s),p(s)]_{0\leq s\leq t}$
\algn{\eqnlab{vft}
	V(m,t)  =  \int_0^t \ed s \left[ p \dot q  - \msc{H}(q,p) \right] + \veq[q(0)]\,.
}
Because of the parity symmetry $m\to-m$ of the problem, the dynamical Hamiltonian $\msc{H}$ in \eqnref{hamiltonian} is invariant under inversion $\msc{H}(q,p) = \msc{H}(-q,-p)$ of $q$ and $p$. Furthermore, $\msc{H}$ satisfies a shift-inversion symmetry in $p$, with respect to the equilibrium rate function $\veqq$ at inverse temperature $\bq$,
\algn{\eqnlab{shiftinv}
	\msc{H}(q,p) = \msc{H}\left[q,-p + \dd{m}\veqq(q)\right]\,,
}
which follows from detailed balance~\cite{Bou20}. For $p=0$, \eqnref{shiftinv} directly yields $\msc{H}(q,0) = \msc{H}[q,\ed\veqq(q)/\ed m] = 0$, which implies, together with~\eqnref{hj}, that the equilibrium rate function $\veqq$ at quenched inverse temperature $\beta_\text{q}$ is invariant under time evolution.

Each pair $[q(s),p(s)]_{0\leq s\leq t}$ of characteristics is associated with a fluctuation path that realises $q(t)=m$ for given $m$ and $t$. While the coordinate $q(s)$ corresponds to the change in the magnetisation $m$ as function of time, the conjugate coordinate $p(s)$ quantifies, roughly speaking, the fluctuations of the environment required to realise $q(t)=m$. In particular, characteristic pairs with $p(s)=0$ correspond to the typical macroscopic relaxation dynamics $\dot q(s) = 2[w_+(q) - w_-(q)]$, $\dot p(s) = 0$, i.e., the dynamics of the minima of $V(m,t)$.

When there are multiple pairs $[q(s),p(s)]_{0\leq s\leq t}$ of characteristics that solve \eqnref{heom} for same boundary conditions \eqnref{bound}, one must pick the pair that minimises $V(m,t)$, since the probabilities of all other solutions are exponentially suppressed~\cite{Mei22a}. The minimising characteristic $q(s)$ represents the most probable way to realise the magnetisation $q(t)=m$ at time $t$, called the \textit{optimal fluctuation} for the event.

\subsubsection{Finite-time dynamical phase transition in magnetisation}\seclab{ftdptmag}
A sudden switch in the optimal fluctuation that realises vanishing magnetisation $m=0$ at the critical time $t_\text{c}$ gives rise to a finite-time dynamical phase transition associated with the magnetisation $m$ in the Curie-Weiss model~\cite{Mei22a,Ent10,Erm10}. This switch can be visualised using the phase portrait of the Hamiltonian dynamics \eqnref{heom}, depicted in Figure~\figref{pp_magnetisation}(a).

The dynamics~\eqnref{heom} occurs along the level lines (shown as the brown lines) of $\msc{H}$ and has a single fixed point (white bullet) at
\algn{\eqnlab{fp}
	q_\text{FP}=p_\text{FP}=0\,,
}
where $\dot q= \dot p = 0$. For infinite observation times ($t\to\infty$), trajectories may approach the fixed point asymptotically along the stable manifold
\algn{\eqnlab{pst}
	p_\text{s}(q) = 0\,,
}
shown as the black line with inwards-pointing arrows in \Figref{pp_magnetisation}(a). Asymptotic escape away from the fixed point, by contrast, must occur along the unstable manifold
\algn{\eqnlab{punst}
	p_\text{u}(q) = \dd{m}\msc{V}_\text{q}^\text{eq}(q)=-\bq q + \frac12[\log(1+q) - \log(1-q)]\,,
}
shown as black line with outwards-pointing arrows.

At small observation times $t\ll\tau$, the optimal fluctuations that realise vanishing magnetisation $q(t)=m=0$ are given by the inactivity solution $q(s)=p(s)=0$ which resides at the fixed point~\eqnref{fp} of the dynamics. However, when $t$ exceeds the critical time $\tc$, which is a function of the parameters of the quench, see \eqnref{tcm} in \Secref{crittime}, non-trivial optimal fluctuations [coloured lines in \Figref{pp_magnetisation}(a)] occur. For times slightly above $\tc$, they remain close to the fixed point, but depart from it more and more as $t$ increases further. For long observation times $t\gg\tau$, the initial points of the optimal fluctuation approach the minima (black bullets) of the initial equilibrium rate function $\veq$. As $t\to\infty$, the dynamics occurs on the stable manifold~\eqnref{pst}, where the fixed point is approached exponentially slowly.
 
The switch of the optimal fluctuation at $t=\tc$ from the inactivity solution for $t\leq\tc$ to non-trivial trajectories for $t>\tc$ gives rise to a finite-time cusp in $V(m,t)$ at $m=0$ for $t>\tc$, shown in \Figref{pp_magnetisation}(b). In close analogy with the equilibrium transition of the model, this cusp can be interpreted as a continuous, finite-time dynamical phase transition~\cite{Mei22a}. The order parameter for the transition, the dynamical analogue for $\bar m(\beta)$ at equilibrium, is given by $q(0)=m_0(m,t)$, the most-likely initial magnetisation for given $m$ and $t$~\cite{Mei22a}. Figure~\figref{pp_magnetisation}(c) shows how $m_0(0,t)$ transitions from zero to a finite value at the critical time $t_\text{c}$, in direct analogy with $\bar m(\beta)$ at equilibrium [\Figref{phase_diag}(b)].

\begin{figure}
	\centering
	\includegraphics[width=9cm]{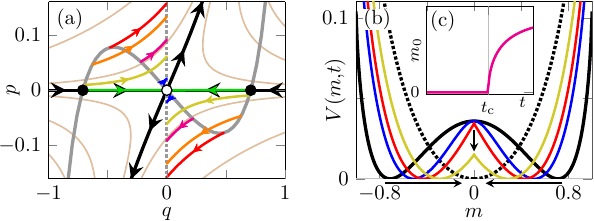}
	\caption{(a) Phase portrait of the Hamilton equations \eqnref{heom} featuring the fixed point (white bullet), the stable and unstable manifolds (black lines), and the level lines of $\msc{H}$ (light brown lines). Gray solid and dotted lines show the initial and final conditions in \eqnref{bound}, respectively. Coloured lines show optimal fluctuations for $t>t_\text{c}\approx 0.3466\tau$, with arrows indicating the evolution in time, for $t/\tau=0.55$ (blue), $0.57$ (magenta), $0.7$ (red), $1$ (orange), $1.75$ (yellow), and $5$ (green). (b) Rate function $V(m,t)$ after disordering quench. The initial equilibrium rate function $\veq$ (solid black line) evolves into $V(m,t)$ at $t/\tau=0.5$ (blue), $0.8$ (red), and $1.5$ (yellow), toward the final equilibrium rate function $\veqq$ (dotted line). Arrows indicate the time evolution. (c) Dynamical order parameter $m_0(m,t)$ as function of $t$ at $m=0$.}\figlab{pp_magnetisation}
\end{figure}
\section{Extended Hamiltonian method for observables}\seclab{thermoobs}
To study the finite-time statistics of observables other than the magnetisation $m$ in the thermodynamic limit, we must generalise the Hamiltonian method described in \Secref{ham}. Although we focus on thermodynamic observables in the Curie-Weiss model, the theory we develop here is generally applicable to systems with well-defined thermodynamic or weak-noise limits and for a larger class of observables (specified below). Every observable in the extended Hamiltonian method is connected to its own dynamical Landau potential. This potential is a powerful tool for the identification and the classification of dynamical phases and phase transitions, associated with a given observable.
\subsection{Thermodynamic observables}
The thermodynamic observables of the Curie-Weiss model and their statistics are the main objects of our study. In order to define them, we use the framework of stochastic thermodynamics~\cite{Sei12,Bro15,Pel21} to endow the stochastic dynamics of the model with a thermodynamic interpretation. This allows us to identify fluctuating thermodynamic observables at the microscale, that are consistent with macroscopic thermodynamics in the limit of large system size.

Because the energy of the system does not change during the temperature quench, the work done on the system vanishes. Hence, the thermodynamics of the quench is determined by the statistics of the heat per spin $\msc{Q}$, released by the system into the environment, and the total entropy $\Sigma$ produced (per spin) in the relaxation process.

The dimensionless released heat \msc{Q} is given by the negative energy change of the system, multiplied by $\bq$:
\algn{\eqnlab{heat}
	\msc{Q}(m,m')  = -\bq[\msc{E}(m) - \msc{E}(m')]\,.
}
Here, $m'$ and $m$ denote two magnetisations before and after the quench, respectively. Both $m'$ and $m$ are random variables which depend on the thermal fluctuations of the heat baths before and after the quench.  In our dimensionless formulation, $\msc{Q}$ is equal to the change in environment entropy per spin, $\msc{Q}=\Sigma^\text{env}$.

The probability distribution $P(\msc{Q},t)$ of $\msc{Q}$ is constrained by a detailed fluctuation relation~\cite{Eva02,Sei12}
\algn{\eqnlab{detfluctrel}
	\frac{P(\msc{Q},t)}{P(-\msc{Q},t)} = \exp\left[-N(\beta/\bq-1)\msc{Q}\right]\,,
}
which relates the negative and positive branches of $P(\msc{Q},t)$. We prove \eqnref{detfluctrel} for the present setup in \secref{detfluctrel}.

The total entropy production $\Sigma$ is the sum of the entropy changes of the environment $\Sigma^\text{env}$ and of the system $\Sigma^\text{sys}$:
\algn{\eqnlab{sig}
	\Sigma(m,m',t) = \Sigma^\text{sys}(m,m',t) + \Sigma^\text{env}(m,m')\,.
}
The change in system entropy $\Sigma^\text{sys}$, in turn, decomposes further into
\algn{\eqnlab{sigsys}
	\Sigma^\text{sys}(m,m',t) = V(m,t) - \veq(m') + \sint(m)-\sint(m')\,.
}
Here, the last two terms represent the change in internal entropy~\eqnref{sint} of the spin system. The first two terms constitute the change in state entropy~\cite{Sei05}, related to the changing probability distribution during the relaxation after the quench. Using \eqnref{heat}, \eqnref{sigsys} and \eqnref{freeen}, we conveniently rewrite \eqnref{sig} as
\algn{\eqnlab{entropy}
	\Sigma(m,m',t)=V(m,t) - \veq(m')-\bq[\msc{F}(m)-\msc{F}(m')]\,.
}
In this formulation, the total entropy production is the sum of the negative change in free energy density (last two terms) and the change in state entropy at time $t$ (first two terms). Note that due to the latter contribution, $\Sigma(m,m',t)$ depends explicitly on time, while $\msc{Q}(m,m')$ does not.
\subsection{Statistics of thermodynamic observables}
Because of the instantaneous nature of the temperature quench $\beta\to\bq$ at $t=0$, the observables $\msc{Q}$ and $\Sigma$ can be written as differences of state functions. More precisely, $\msc{Q}$ and $\Sigma$ in \eqnref{heat} and \eqnref{entropy} are differences of $-\bq\msc{E}$ and $V-\bq\msc{F}$, respectively, evaluated at $(m,t)$ and $(m',0)$. Observables of this kind depend only on the initial and final states, $m'$ and $m$, and on time $t$, but are otherwise independent of the specific path $m(s)_{0\leq s\leq t}$ taken by the dynamics.

Based on this observation, we now develop a general theory for the large-deviation statistics of such observables that applies to systems subject to quenches of the temperature or of other external parameters. This includes, but is not limited to, the thermodynamic observables $\msc{Q}$ and $\Sigma$.

We define the moment-generating function $G(k,t)$ of an intensive state-variable difference 
\algn{\eqnlab{statevar}
	\Delta \msc{A} = \msc{A}(m,t) - \msc{A}(m',0)\,,
}
by
\algn{\eqnlab{mgf}
	G(k,t) = \langle \ee^{N k \Delta\msc{A}}\rangle\,.
}
Note again that the explicit time dependence of $\msc{A}$ in \eqnref{statevar} is absent for $\msc{Q}$ but present in $\Sigma$. Conditioning \eqnref{mgf} on the initial and final magnetisations $m'$ and $m$, we write
\algn{\eqnlab{mgf0}
	G(k,t) = \sum_{m,m'}\langle \ee^{N k \Delta\msc{A}}|m,t;m',0\rangle P(m,t;m',0)\,,
}
where $P(m,t;m',0)$ denotes the joint probability of observing $m$ at time $t$ and $m'$ at vanishing initial time. Because the observable $\Delta\msc{A}$ in \eqnref{mgf0} depends only on $m$, $m'$, and $t$, the conditioning renders $\Delta\msc{A}$ deterministic, so that  $\langle \ee^{N k \Delta\msc{A}}|m,t;m',0\rangle = \ee^{N k \Delta\msc{A}}$. Furthermore, we write the joint probability in \eqnref{mgf0} as $P(m,t;m',0) = P_\text{q}(m,t|m',0)\peq(m')$, where $P_\text{q}(m,t|m',0)$ denotes the probability of observing magnetisation $m$ at time $t$ \textit{conditional} on starting with $m'$ at time $t=0$. The subscript $q$ emphasises that the dynamics is due to the heat bath at quenched inverse temperature $\bq$. After these manipulations, \eqnref{mgf0} reads
\algn{\eqnlab{mgf1}
	G(k,t) = \sum_{m,m'}\ee^{N k [\msc{A}(m,t) - \msc{A}(m',0)]} P_\text{q}(m,t|m',0)\peq(m')\,,
}
We now define the ``$k$-tilted'' initial probability distribution
\algn{\eqnlab{pkinit}
	P_k(m',0) \equiv Z_k^{-1}\ee^{-N k\msc{A}(m',0)}\peq(m')\,,
}
with $Z_k$ obtained from normalisation. Summing over $m'$ in \eqnref{mgf1} we then arrive at
\algn{\eqnlab{mgf2}
	G(k,t) = Z_k\langle \ee^{N k \msc{A}}\rangle_k\,,
}
where $\langle \ldots \rangle_k$ denotes the average with respect to the time-evolved, $k$-tilted distribution
\algn{\eqnlab{pkdist}
	P_k(m,t) = \sum_{m'}P_\text{q}(m,t|m',0)P_k(m',0)\,.
}
Equation~\eqnref{mgf2} has the advantage that the observable $\msc{A}(m,t)$ is independent of the initial state $m'$, in distinction to $\Delta\msc{A}$ in \eqnref{mgf}, which depends on both $m'$ and $m$.
\subsection{Thermodynamic limit and Landau potential}
In the thermodynamic limit $N\to\infty$, the probability distribution of an intensive observable $\Delta\msc{A}$ typically takes a large-deviation form~\cite{Ell07,Hol08,Tou09}
\algn{\eqnlab{rateda}
	P(\Delta\msc{A},t) \smilefrown \ee^{- N I(\Delta\msc{A},t)}\,,
}
with non-negative rate function $I(\Delta\msc{A},t)\geq0$, analogous to $V(m,t)$ in \eqnref{ratem}. The location of the vanishing minimum of $I(\Delta\msc{A},t)$ is given by the typical, most probable value of $ \Delta\msc{A}$, which coincides with its mean $\langle\Delta\msc{A}\rangle$, i.e., $I(\langle \Delta\msc{A}\rangle,t)=0$, in the thermodynamic limit. Away from this minimum, the rate function quantifies the exponentially suppressed probabilities of deviations from the typical behaviour, thus generalising the central-limit theorem~\cite{Ell07,Hol08,Tou09}. In other words, the rate function $I(\Delta\msc{A},t)$ provides us with the time-dependent statistics of $\Delta\msc{A}$, to leading exponential order in the thermodynamic limit.

It is convenient to use the scaled cumulant-generating function
\algn{\eqnlab{scgf}
	\Lambda(k,t)\equiv\lim_{N\to\infty}\frac1N\ln G(k,t)\,,
}
to obtain $I(\Delta\msc{A},t)$ by Legendre transform~\cite{Ell07,Hol08,Tou09}
\algn{\eqnlab{legendre}
	I(\Delta\msc{A},t) = \max_{k}\{ k\Delta \msc{A} - \Lambda(k,t)		\}\,.
}
The scaled cumulants of $\Delta\msc{A}$ are given by the derivatives of $\Lambda(k,t)$, evaluated at $k=0$. In particular, the mean is given by the slope at $k=0$,
\algn{\eqnlab{meanA}
	\langle\Delta\msc{A}\rangle = \partial_k\Lambda(0,t)\,.
}
For $\Delta\msc{A} = \msc{Q}$, the detailed fluctuation relation \eqnref{detfluctrel} implies a symmetry for $\Lambda(k,t)$ about the inflection point $k_0=(\beta/\bq-1)/2$:
\algn{\eqnlab{detfluctrellam}
	\Lambda(k+k_0,t)=\Lambda(-k+k_0,t)\,.
}
In order to derive an expression for $\Lambda(k,t)$, we take the thermodynamic limit of \eqnref{mgf2} using the large-deviation form 
\algn{
	 P_k(m,t)\smilefrown\ee^{-NV_k(m,t)}
}
for the $k$-tilted probability distribution, with $k$-tilted rate function $V_k(m,t)$. In the limit $N\to\infty$, the sums in \eqnref{mgf2} and \eqnref{pkdist} turn into integrals that we evaluate by a saddle-point approximation. We collect the exponential terms and substitute them into \eqnref{scgf}, which yields 
\algn{\eqnlab{scgf2}
	\Lambda(k,t) = -\min_m\left\{W_k(m,t)\right\}\,,
}	
where
\algn{\eqnlab{wk}
	W_k(m,t) = -k\msc{A}(m,t) + V_k(m,t) -\zeta_k\,.
}
Equation~\eqnref{scgf2} expresses $\Lambda(k,t)$ as the negative minimum of the potential function $W_k(m,t)$. We show in \Secref{ftdpt} that $W_k(m,t)$ takes the role of a dynamical Landau potential. The $k$-dependent constant 
\algn{
	\zeta_k=\lim_{N\to\infty}\frac1N\ln Z_k\,,
}
in \eqnref{wk} originates from the normalisation of the tilted rate function $V_k(m,t)$, but it cancels in the expressions of $W_k(m,t)$ and $\Lambda(k,t)$, as we shall see in \eqnref{wk2} and \eqnref{lamk} below.

From \eqnref{pkdist}, we observe that the $k$-tilted rate function $V_k(m,t)$ obeys, up to different boundary conditions, the same Hamilton-Jacobi equation~\eqnref{hj} as the ``untilted'' magnetisation rate function $V(m,t)=V_{k=0}(m,t)$, i.e.,
\algn{\eqnlab{hj2}
	0=\partial_t V_k(m,t) + \msc{H}[m,\partial_m V_k(m,t)]\,.
}
The initial condition for $V_k(m,t)$ follows from the large-deviation form of $P_k(m,0)$ in \eqnref{pkinit}:
\algn{\eqnlab{vbound}
	V_k(m,0)= k\msc{A}(m,0) + \veq(m) + \zeta_k\,.
}
As a consequence, the time evolution of $W_k(m,t)$ is also dictated by \eqnref{hj2} through \eqnref{wk}. 

Before we proceed, we find it instructive to compute $I(\Delta\msc{A},t)$ at $t=0$ with what we have derived so far. Substituting the boundary condition~\eqnref{vbound} into \eqnref{wk} at $t=0$, we find $W_k(m,0)=\veq(m)$. Equation~\eqnref{scgf2} then gives $\Lambda(k,0)=0$ for all $k$. Performing the Legendre transform~\eqnref{legendre}, we find $P(\Delta\msc{A},0)=\delta(\Delta\msc{A})$, where $\delta(x)$ denotes the Dirac delta function. This shows that the state-difference observable $\Delta\msc{A}$ in \eqnref{statevar} is initially zero with unit probability for all observables $\msc{A}$, as expected.
\subsection{Post-quench dynamics of \texorpdfstring{$W_k(m,t)$}{Wk(m,t)}}\seclab{wkdyn}
At finite time $t>0$ after the quench, $V_k(m,t)$ is the solution of the Hamilton-Jacobi equation~\eqnref{hj2} with initial condition~\eqnref{vbound}. Equation~\eqnref{hj2} is solved by a $k$-dependent family of characteristics $[q_k(s),p_k(s)]_{0\leq s\leq t}$ that are solutions to the Hamilton equations~\eqnref{heom} with $k$- and $m$-dependent boundary conditions:
\algn{\eqnlab{pbound}
	p_k(0) = k\partial_m\msc{A}[q_k(0),0] +\tfrac{\ed}{\ed m}\veq[q_k(0)]\,,\quad q_k(t) =& m\,.
}
An expression for $V_k(m,t)$ is then given by the $k$-tilted analogue of \eqnref{vft},
\algn{\eqnlab{vkft}
	V_k(m,t) = \int_0^t \ed s \left[ p_k \dot q_k  - \msc{H}(q_k,p_k) \right] + V_k[q_k(0),0]\,.
}
Using \eqnref{wk}, we can now write $W_k(m,t)$ in terms of the characteristics $[q_k(s),p_k(s)]_{0\leq s\leq t}$ as
\algn{\eqnlab{wk2}
	W_k(m,t) = \int_0^t \ed s \left[ p_k \dot q_k  - \msc{H}(q_k,p_k) \right] + \veq[q_k(0)] -k\{\msc{A}[q_k(t),t]-\msc{A}[q_k(0),0]\}\,.
}
Hence, using~\eqnref{wk2} one computes $W_k(m,t)$ from the solutions $[q_k(s),p_k(s)]_{0\leq s\leq t}$ of \eqnref{heom} with boundary conditions \eqnref{pbound}, for given $m$, $k$ and $t$. In order to obtain $\Lambda(k,t)$ from $W_k(m,t)$ according to \eqnref{scgf2}, we must then take an additional, error-prone minimum over $m$. Hence, although \eqnref{wk2} is useful for computing the Landau potential $W_k(m,t)$, it is not the ideal starting point for the evaluation of $\Lambda(k,t)$. In the next section, we derive a method for computing $\Lambda(k,t)$, that avoids evaluating $W_k(m,t)$ for all $m$ and only requires a one-dimensional $k$ grid.
\subsection{Post-quench dynamics of \texorpdfstring{$\Lambda(k,t)$}{L(k,t)}}
Equation~\eqnref{wk2} provides an expression for $W_k(m,t)$ for given $m$ and $t$. The scaled cumulant-generating function $\Lambda(k,t)$ in \eqnref{scgf2}, however, requires only the value $W_k(m^*_k,t)$ where $W_k(m,t)$ acquires its minimum, i.e., $\partial_mW_k(m^*_k,t)=0$. The minimum value $W_k(m^*_k,t)$ can be obtained directly by imposing $\partial_mW_k(m^*_k,t)=0$ as a boundary condition, in addition to \eqnref{vbound}. Written as a condition for $V_k(m,t)$, one finds from \eqnref{wk}:
\algn{\eqnlab{vbound2}
	\partial_mV_k(m_k^*,t) = k\partial_m\msc{A}(m_k^*,t)
}
at time $t$. This boundary condition leads to yet another family of characteristics $[q^*_k(s),p^*_k(s)]_{0\leq s\leq t}$, which, again, obey the Hamilton equations~\eqnref{heom}, but now with $m$-independent boundary conditions
\sbeqs{\eqnlab{psbound}
\algn{
	p^*_k(0) =& k\partial_m\msc{A}[q^*_k(0),0] +\tfrac{\ed}{\ed m}\veq[q^*_k(0)]\,,	\eqnlab{psbound1}\\
	p^*_k(t) =& k\partial_m\msc{A}[q_k^*(t),t]\,.	\eqnlab{psbound2}
}
}
These boundary conditions ensure that $q_k^*(t)=m_k^*$ is an extremum of $W_k(m,t)$. Using this set of characteristics $[q^*_k(s),p^*_k(s)]_{0\leq s\leq t}$, we directly express $\Lambda(k,t)$ as
\algn{\eqnlab{lamk}
	\Lambda(k,t) = -\int_0^t \ed s \left[ p^*_k \dot q^*_k  - \msc{H}(q^*_k,p^*_k) \right] - \veq[q^*_k(0)]+k\{\msc{A}[q^*_k(t),t]-\msc{A}[q^*_k(0),0]\}\,.
}
In the particular case $\Delta\msc{A}=\msc{Q}$, that we study in detail in \Secref{ftdpt}, the shift-inversion symmetry~\eqnref{shiftinv} of $\msc{H}$, combined with the boundary conditions~\eqnref{psbound}, implies a time-reversal symmetry for characteristics below and above $k_0$,
\algn{\eqnlab{timerev}
	q^*_{k+k_0}(s) = q^*_{-k+k_0}(t-s)\,,\qquad p^*_{k+k_0}(s) = -p^*_{-k+k_0}(t-s) + \veqq[q^*_{-k+k_0}(t-s)]\,.
}
This is a generalisation of the detailed fluctuation relation~\eqnref{detfluctrellam} to the level of optimal fluctuations, which can be seen by recovering~\eqnref{detfluctrellam} from \eqnref{lamk} and \eqnref{timerev}.

Depending on whether we intend to compute $W_k(m,t)$ or $\Lambda(k,t)$, we solve the Hamilton equations~\eqnref{heom} with either boundary conditions, \eqnref{pbound} or \eqnref{psbound}, by a shooting method~\cite{Mei22a}. This returns families of characteristics on a grid of $k$ (and $m$) values, enabling us to evaluate either $W_k(m,t)$ in \eqnref{wk}, or $\Lambda(k,t)$ in \eqnref{lamk} on this grid.
\subsection{Long-time limit}
In the long-time limit, the evaluation of $\Lambda(k,t)$ simplifies considerably. This is seen most directly by taking the long-time limit of $G(k,t)$ in \eqnref{mgf1}, where the conditional probability converges to the equilibrium probability distribution $\peqq$ at the quenched inverse temperature \bq, $\lim_{t\to\infty}P_\text{q}(m,t|m',0)=\peqq(m)$. We can then write the limit $G_\infty(k) \equiv\lim_{t\to\infty}G(k,t)$ as
\algn{
	G_\infty(k) = \langle \ee^{Nk \msc{A}_\infty(m)}\rangle^\text{eq}_q \langle \ee^{-Nk \msc{A}(m',0)}\rangle^\text{eq}\,,
}
where $\msc{A}_\infty(m)\equiv\lim_{t\to\infty}\msc{A}(m,t)$; $\langle\ldots\rangle^\text{eq}$ and $\langle\ldots\rangle^\text{eq}_\text{q}$ denote averages with respect to the equilibrium distributions $\peq$ and $\peqq$, respectively. Taking the thermodynamic limit, we use the large-deviation forms of these distributions and evaluate the integrals over $m$ and $m'$ in the saddle-point approximation. This leads us to an expression for $\Lambda_\infty(k)\equiv\lim_{t\to\infty}\Lambda(k,t)$ in terms of a maximisation over initial and final states $m$ and $m'$, given by
\algn{\eqnlab{laminf}
	\Lambda_\infty(k) = \max_m\{k\msc{A}_\infty(m) - \veq_q(m)	\}+ \max_{m'}\{-k\msc{A}(m',0) - \veq(m')	\}\,.
}
To connect this to our previous results, we write \eqnref{laminf} as a function of the initial and final points, $q_k^*(0)$ and $q_k^*(\infty)$, of an infinite-time optimal fluctuation:
\algn{\eqnlab{qkinf}
	\Lambda_\infty(k) =	k\msc{A}_\infty[q_k^*(\infty)] - \veqq[q_k^*(\infty)]-k\msc{A}[q_k^*(0),0] - \veq[q_k^*(0)]\,.
}
In order for the optimal fluctuation to fulfil the boundary conditions \eqnref{psbound} in infinite time, it must initiate on the stable manifold~\eqnref{pst}, pass through the fixed point at $(q_\text{FP},p_\text{FP})=(0,0)$, and either stay there [when $q_{k}^*(\infty)=0$], or end on the unstable manifold~\eqnref{punst} [when $q_{k}^*(\infty)\neq0$]. Combining \eqnref{pst} with \eqnref{psbound1} at $q=q_k^*(0)$ and \eqnref{punst} with \eqnref{psbound2} at $q=q_k^*(\infty)$, we find that the initial and end points must satisfy
\sbeqs{\eqnlab{psboundinf}
	\algn{
		0 =& k\partial_m \msc{A}[q_k^*(0),0]+\dd{m}\veq[q_k^*(0)]\,,\\
		0 =& k\partial_m \msc{A}_\infty[q_k^*(\infty)]-\dd{m}\veqq[q_k^*(\infty)]\,.
	}
}
In case there are several solutions to \eqnref{psboundinf}, we must pick the combination of $q_k^*(0)$ and $q_k^*(\infty)$ for which the right-hand side of \eqnref{qkinf} takes its maximum value. This approach leads to explicit expressions for the scaled cumulant-generating function $\Lambda_\infty(k)$ and for the initial and final points of $q_k^*(s)_{0\leq s\leq\infty}$ in the infinite-time limit.
\subsection{Rate function}
Finally, we compute the rate function $I(\Delta\msc{A},t)$ from $\Lambda(k,t)$ using the Legendre transform in \eqnref{legendre}. To this end, we evaluate
\algn{\eqnlab{legda}
	\Delta\msc{A}(k^*) = \partial_k\Lambda(k^*,t) = \msc{A}[q^*_{k^*}(t),t]-\msc{A}[q^*_{k^*}(0),0]\,,
}
which gives an implicit equation for the value $k^*$, where the right-hand side of \eqnref{legendre} acquires its maximum. The second equality in \eqnref{legda} follows by taking a $k$ derivative of \eqnref{lamk}:
\algn{\eqnlab{lamkder}
	\partial_k\Lambda(k,t) = \msc{A}[q^*_{k}(t),t]-\msc{A}[q^*_{k}(0),0]+\int_0^t \ed s   \left[\frac{\delta \Lambda}{\delta p^*_k} \frac{\partial p^*_k}{\partial k} +  \frac{\delta \Lambda}{\delta q^*_k} \frac{\partial q^*_k}{\partial k}\right]\,.
}
The integral in \eqnref{lamkder} vanishes due to the variational principle $\delta\Lambda=0$, which implies $\delta\Lambda/\delta q_k^*=\delta\Lambda/\delta p_k^*=0$, see \secref{variation}. 

By inverting \eqnref{legda} we obtain the function $k^*(\Delta\msc{A},t)$ and express \eqnref{legendre} as
\algn{\eqnlab{ratefda}
	I(\Delta\msc{A},t) = k^*(\Delta\msc{A},t)\Delta\msc{A}-\Lambda[k^*(\Delta\msc{A},t),t]\,.
}
Equation~\eqnref{ratefda} is the sought-after expression for the rate function $I(\Delta\msc{A},t)$ in \eqnref{rateda}, that completes the extended Hamiltonian method for the calculation of the finite-time statistics of observables in the thermodynamics limit.

To summarise, while the standard Hamiltonian method, outlined in \Secref{ham}, only yields the statistics of the state (magnetisation $m$ for the Curie-Weiss model), the extended Hamiltonian method developed here represents a significant generalisation. It computes the finite-time large-deviations of arbitrary $\Delta\msc{A}$ in the thermodynamic limit, whenever $\Delta\msc{A}$ can be written as a difference of state functions. For the post-quench dynamics we analyse here, the thermodynamic observables $\msc{Q}$ and $\Sigma$ fall into this category. Furthermore, the method automatically provides the dynamical Landau potential $W_k(m,t)$ given in~\eqnref{wk2}, a powerful tool of the identification and classification of finite-time dynamical phase transitions.

We note that since our method uses generating functions and the Legendre transform, $I(\Delta\msc{A},t)$ obtained from \eqnref{ratefda} is always convex. In cases where the underlying rate function has a non-convex part, such as, e.g., $V(m,t)$ in \Figref{pp_magnetisation}(b), \eqnref{ratefda} returns its convex hull, and the information stored in the non-convex part is lost. Non-convex parts in rate functions are signalled by non-differentiable points~\cite{Tou09} in the scaled-cumulant generating function $\Lambda(k,t)$ in \eqnref{scgf}. As a simple but non-trivial example of $\Delta\msc{A}$ where this non-convexity matters, one may consider the magnetisation itself, $\Delta\msc{A} = m$, so that $\msc{A}(m,t) = m$ and $\msc{A}(m',0) = 0$. In this case, the scaled cumulant generating function $\Lambda(k,t)$ has a sharp kink at $k=0$, because the magnetisation rate function $V(m,t)$ is non-convex between its minima for all finite times, see \Figref{pp_magnetisation}(b). The rate function $I(m,t)$ obtained from \eqnref{ratefda} then gives the convex hull of $V(m,t)$, which is flat around the origin and, in particular, misses the kink of $V(m,t)$ at $m=0$.

Fortunately, $\Lambda(k,t)$ is oftentimes differentiable in its domain, and \eqnref{ratefda} returns the exact rate function. This is the case for thermodynamic observables of the Curie-Weiss model, so that non-convexity is of no concern here.
\section{Finite-time dynamical phase transition}\seclab{ftdpt}
We now apply the theory developed in the previous section to the time-dependent large-deviation statistics of $\msc{Q}$ in the Curie-Weiss model. The total entropy production $\Sigma$, see~\eqnref{entropy}, can be analysed in an analogous way.

In order to compute $\Lambda(k,t)$ for $\msc{Q}$ at finite time $t$, we solve the Hamiltonian equations~\eqnref{heom} with boundary conditions \eqnref{psbound} to obtain $[q^*_k(s),p^*_k(s)]_{0\leq s\leq t}$ for a grid of $k$ values, and evaluate \eqnref{lamk} on this grid.

\begin{figure}
	\centering
	\includegraphics[width=9cm]{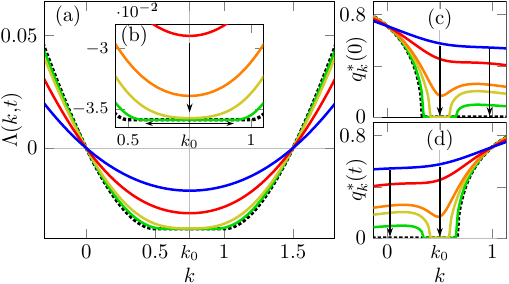}
	\caption{Post-quench evolution of scaled-cumulant generating function $\Lambda(k,t)$ for $\beta=5/(4J)$ and $\bq=1/(2J)$, with critical time $\tcenv/\tau = \ln(3)\approx1.0986$ computed in \Secref{crittime}, see \eqnref{tcenv}. Arrows indicate changes in time. (a) $\Lambda(k,t)$ for $t/\tau = 0.25$ (blue), $0.5$ (red), $1$ (yellow), $1.5$ (green), and $\infty$ (dotted). (b) Magnified view of the flat region in \Figref{scgf_senv}(a), including $\Lambda(k,t)$ for $t/\tau=0.75$ (orange). (c)--(d) Initial and final magnetisations $q^*_k(0)$ and $q^*_k(t)$ for $t/\tau=0.25$ (blue), $0.5$ (red), $1$ (orange), $1.25$ (yellow), $2$ (green), and $\infty$ [dotted, obtained from \eqnref{psboundinf}].}\figlab{scgf_senv}
\end{figure}

Figure~\figref{scgf_senv}(a) shows $\Lambda(k,t)$ at different times $t>0$ after the quench. We observe that $\Lambda(k,t)$ has an initially parabolic shape, but develops a flat region around its inflection point $k_0=(\beta/\bq-1)/2$ at a finite, critical time $\tcenv$. This critical time is of the order of the microscopic relaxation time $\tau$ and is specified in \Secref{crittime}, see \eqnref{tcenv}. For $t/\tau>1$, we observe quick convergence towards the long-time limit $\Lambda_\infty(k)$ (dotted line), obtained from \eqnref{laminf}. Figure~\figref{scgf_senv}(b) shows a magnification of the flat region in \Figref{scgf_senv}(a); the arrows indicate the evolution in time. 

Figures~\figref{scgf_senv}(c) and \figref{scgf_senv}(d) show the optimal initial and final magnetisations $q^*_k(0)$ and $q^*_k(t)$ at different times. Note that there exists an equivalent, negative pair $-q^*_k(0)$ and $-q^*_k(t)$, due to the parity symmetry $m\to-m$ of the problem. Furthermore, the time-reversal symmetry~\eqnref{timerev}, evaluated at $s=0$, relates the initial and end points of the optimal fluctuations in \Figsref{scgf_senv}(c) and \figref{scgf_senv}(d).

At short times when $\Lambda(k,t)$ is parabolic, both $q^*_k(0)$ and $q^*_k(t)$ are finite. For $t>\tcenv$, by contrast, $q^*_k(0)=q^*_k(t)=0$ in the finite $k$ region around $k_0$ where $\Lambda(k,t)$ is flat. This indicates that the inactivity solution $q^*_k(s)=p^*_k(s)=0$ is the optimal fluctuation within this $k$ interval, leading to the flat region in $\Lambda(k,t)$. A more detailed analysis of the optimal fluctuations, conducted in \Secref{optfluct}, provides an intuitive explanation for this, based on the constrains on the optimal fluctuations imposed at $\msc{Q}=0$. Substituting the inactivity solution into \eqnref{lamk}, we find the constant value $\Lambda(k,t) = -\veq(0) = \ln(2) + \beta\msc{\bar F} (\beta)\approx -0.0359$ for $t>t^\msc{Q}_\text{c}$ within the flat region.

For longer times, we observe convergence, indicated by the black arrows, of the initial and end points $q^*_k(0)$ and $q^*_k(t)$ toward the asymptotic, long-time solution (dotted lines), obtained from \eqnref{psboundinf}. From \eqnref{laminf} we obtain the asymptotic boundaries $k_\text{min}$ and $k_\text{max}$ of the flat interval as
\algn{\eqnlab{boundaries}
	k_\text{min} = \frac{\beta-\bc}{\bq}\,,\qquad k_\text{max} = \frac{\bc-\bq}{\bq}\,,
}
which evaluate to $k_\text{min} = 1/2$ and $k_\text{max} = 1$ for the parameters of \Figref{scgf_senv}.

\begin{figure}
	\centering
	\includegraphics[width=9cm]{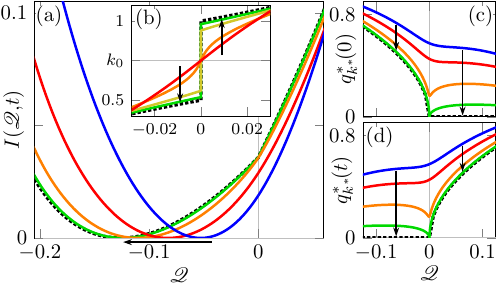}
	\caption{Post-quench evolution of rate function $I(\msc{Q},t)$ for $\beta=5/(4J)$ and $\bq=1/(2J)$, with critical time $\tcenv/\tau = \ln(3)\approx1.0986$ computed in \Secref{crittime}, see \eqnref{tcenv}. Arrows indicate changes in time. (a) $I(\msc{Q},t)$ for $t/\tau = 0.25$ (blue), $0.5$ (red), $1$ (orange), $2$ (green) and $\infty$ (dotted). (b) Derivative $\partial_\msc{Q}I(\msc{Q},t)$ in small interval around kink, including $\partial_\msc{Q}I(\msc{Q},t)$ for $t/\tau=1.5$ (yellow). (c)--(d) Initial and final magnetisations $q^*_{k^*}(0)$ and $q^*_{k^*}(t)$ for times $t/\tau = 0.25$ (blue), $0.5$ (red), $1$ (orange), $1.5$ (green) and $\infty$ (dotted).}\figlab{rf_senv}
\end{figure}

By the Legendre transform~\eqnref{legendre} of $\Lambda(k,t)$, we obtain the rate function $I(\msc{Q},t)$, shown in \Figref{rf_senv}(a). At the critical time $\tcenv$ when $\Lambda(k,t)$ starts developing the flat region, $I(\msc{Q},t)$ acquires a kink around vanishing $\msc{Q}$. The location $\msc{Q}=0$ of the kink is determined by the vanishing slope of the flat $k$-interval in $\Lambda(k,t)$. At the kink, the derivative $\partial_\msc{Q}I(\msc{Q},t)$ attains a finite jump, centered at $k_0$ [see \Figref{rf_senv}(b)], whose magnitude, in turn, corresponds to the width of the flat $k$-interval in $\Lambda(k,t)$.

The minimum of the rate function represents the typical, average, amount of heat $\langle \msc{Q}\rangle$ released from the system into the environment. As time evolves, $\langle \msc{Q}\rangle$ takes increasingly negative values, because the typical net heat flow occurs from the environment into the spin system, increasing its temperature, $\beta\to\bq$. For $t\gg\tau$, $\langle \msc{Q}\rangle$ settles at a finite value, while the spins equilibrate with the environment. During this process, the value $I(0,t)$ of the rate function at the kink increases, which implies that the event $\msc{Q}=0$ becomes less typical, i.e., less probable, at larger times.

Figures~\figref{rf_senv}(c) and \figref{rf_senv}(d) show the (positive) initial and final magnetisations $q^*_{k^*}(0)$ and $q^*_{k^*}(t)$ as functions of the heat $\msc{Q}$ they generate. As the critical time $\tcenv$ is approached, both $q^*_{k^*}(0)$ and $q^*_{k^*}(t)$ develop a cusp at $\msc{Q}=0$, the location of the kink in $I(\msc{Q},t)$. The cusp is sharp and non-differentiable for $t\geq\tcenv$, and arises because the Legendre transform~\eqnref{legendre} contracts the finite, flat $k$-interval in \Figsref{scgf_senv}(c) and \figref{scgf_senv}(d) where $q^*_k(0)=q^*_k(t)=0$ to the single point $\msc{Q}=0$. Consequently, for $t\geq\tcenv$, $q^*_{k^*}(0)$ and $q^*_{k^*}(t)$ vanish at $\msc{Q}=0$ but are otherwise finite. As function of $\msc{Q}$, the symmetry~\eqnref{timerev} implies, that the optimal fluctuations for $\msc{Q}$ and $-\msc{Q}$ are related by time reversal, so that their initial and end points in \Figsref{rf_senv}(c) and \figref{rf_senv}(d) swap places for $\msc{Q}\to-\msc{Q}$. In the long-time limit, we observe asymptotic convergence towards the long-time limit (dotted lines), obtained from \eqnref{psboundinf}.

In the spirit of the equilibrium analysis of \Secref{equilibrium}, we interpret the development of the flat region in the scaled cumulant-generating function $\Lambda(k,t)$ and the kink in the rate function $I(\msc{Q},t)$ as a finite-time dynamical phase transition. This transition appears similar to the finite-time cusp in $V(m,t)$ discussed in Ref.~\cite{Mei22a} and shown in \Figref{pp_magnetisation}(b), but its properties and origin are different. To proceed, we first identify the optimal, \textit{final} magnetisation $m_t= q^*_k(t)$ as the dynamical order parameter. This is a natural choice, because within the flat region in $\Lambda(k,t)$ and at the kink of $I(\msc{Q},t)$ $q^*_k(t)$ is finite for $t<\tcenv$ and vanishes otherwise, indicating the existence of different dynamical phases. Since $\pm q^*_k(t)$ are the minima of $W_k(m,t)$, see \eqnref{scgf2}, $W_k(m,t)$ takes the role of a dynamical Landau potential, with minima given by $\pm q^*_k(t)$, in close analogy with $\veq(m)$ at equilibrium.

The dynamical phases of the transition in $\msc{Q}$ are associated with the shape of $W_k(m,t)$ in the $t$-$k$ (and $t$-$\msc{Q}$) parameter plane. The number of minima $\pm q^*_k(t)$ suggests two extended phases, shown in \Figref{dyn_phasediag}(a). In the dynamical coexistence (DCE) phase (white region) $W_k(m,t)$ has two minima at $\pm q^*_k(t)$ and the dynamical order parameter $m_t=q^*_k(t)$ is finite. In the dynamical single mode (DSM) phase [lined region in \Figref{dyn_phasediag}(a)] $W_k(m,t)$ has a vanishing unique minimum, so that $m_t=0$. The two phases are separated by a phase boundary (solid line) that emerges from the critical point $(\tcenv,k_0)$ (orange bullet).

\begin{figure}
	\centering
	\includegraphics[width=9cm]{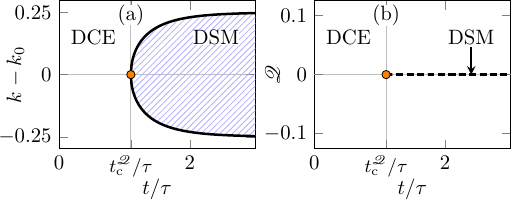}
	\caption{(a) Phase diagram for the finite-time dynamical phase transition of $\msc{Q}$ in the $t$-$k$ plane, featuring the DSM phase (lined) and the DCE phase (white), separated by a phase boundary (black line) that emerges from the critical point $(\tcenv,k_0)$ (orange bullet). The extended DSM phase corresponds to the flat region in $\Lambda(k,t)$. (b) Phase diagram in the $t$-$\msc{Q}$ plane after Legendre transform, with DSM phase contracted to the dashed line.}\figlab{dyn_phasediag}
\end{figure}

Comparing \Figsref{scgf_senv}(c)--(d) with \Figsref{rf_senv}(c)--(d), we observe that due to the nature of the Legendre transform~\eqnref{legendre}, the extended DSM phase in the $t$-$k$ plane contracts to a line at $\msc{Q}=0$, in the $t$-$\msc{Q}$ parameter space. Hence, the phase diagram transforms into a cut plane, with respect to the physical parameters $t$ and $\msc{Q}$. This is shown in \Figref{dyn_phasediag}(b), where the DSM phase is given by the dashed line. 

The cut-plane topology of the phase diagram provides an intuitive explanation of the formation of the kink in $I(\msc{Q},t)$ at $\msc{Q}=0$ for $t>\tcenv$: When $\msc{Q}=0$ is crossed for $t<\tcenv$, i.e., without crossing the DSM phase, the order parameter remains finite and $I(\msc{Q},t)$ is smooth. For $t>\tcenv$, however, $I(\msc{Q},t)$ must cross the DSM phase at $\msc{Q}=0$. At the crossing, $m_t$ becomes zero and bounces back non-differentiably, see the green line in \Figref{rf_senv}(d), resulting in the kink in $I(\msc{Q},t)$.
\subsection{Characterisation of phase transition}\seclab{charphase}
We now give a more detailed characterisation of the dynamical phase transition in terms of the dynamical Landau potential $W_{k}(m,t)$ defined in Sec~\secref{wkdyn}. This allows us to establish the continuous nature of the transition, to obtain an explicit expression for the critical time $\tcenv$ and the critical exponent, and to provide an intuitive explanation for the occurrence of the transition in terms of a switch in the optimal fluctuations.

To get started, we first establish a connection between the locations of the kink in the $k$ and $\msc{Q}$ spaces. We take a $k$ derivative of the detailed fluctuation relation~\eqnref{detfluctrellam}, and evaluate at $k=k_0$:
\algn{
	\partial_k \Lambda(k_0,t) = -	\partial_k \Lambda(k_0,t) = 0\,.
}
By \eqnref{legda}, the derivative at $k_0$ is connected to the value of the observable $\msc{Q}$ generated by the $k$-tilted dynamics as
\algn{
	\partial_k \Lambda(k_0,t) = \msc{Q}(k_0,t) = 0\,.
}
Inverting this equation, we find  $k^*(0,t)=k_0=(\beta/\bq-1)/2$, establishing that $q^*_{k_0}(s)_{0\leq s\leq t}$ is the optimal fluctuation that generates $\msc{Q}=0$ for all times $t$. In particular, this means that we can write the dynamical order parameter $m_t(\msc{Q})$ for $\msc{Q}=0$, as $m_t(0) = q^*_{k_0}(t)$ and that the dynamical Landau potential of the transition at $\msc{Q}=0$ is given by $W_{k_0}(m,t)$.

After identifying $W_{k_0}(m,t)$ as our object of study, we compute it with the method described in \Secref{wkdyn}: We solve the Hamilton equations~\eqnref{heom} with boundary conditions~\eqnref{pbound} to obtain a one-parameter family of characteristics on a fine $m$-grid. From~\eqnref{vkft} we then evaluate $V_{k_0}(m,t)$ on this grid. In the last step, we compute $W_{k_0}(m,t)$ using \eqnref{wk}. 

Figure~\figref{wk0}(a) shows $W_{k_0}(m,t)$ for different $t$ after a quench with the same parameters as in \Figref{scgf_senv} and \Figref{rf_senv}. As expected from the previous discussion, $W_{k_0}(m,t)$ is initially of double-well shape but transitions into a single well at $t=\tcenv\sim\tau$ (coloured lines). At the same time, the dynamical order parameter $m_t$ (bullets), passes from finite to zero. 

Figure~\figref{wk0}(b) shows $W_{k_0}(m,t)$ after a quench with a different set of parameters. In this case, we observe that $W_{k_0}(m,t)$ retains its double-well shape at all times, so that the order parameter $m_t$ remains finite, $m_t>0$. In other words, although the second quench also crosses the phase boundary in \Figref{phase_diag}(a) (orange arrow), it does not induce a finite-time dynamical phase transition for $\msc{Q}$. This shows that the requirement that the quench be disordering, i.e., $\bq<\bc<\beta$, does \textit{not} ensure that the phase transition in $\msc{Q}$ takes place. This is in contrast to the transition in the magnetisation $m$~\cite{Mei22a}, which occurs for \textit{all} disordering quenches.

Furthermore, as shown in the magnified view in \Figref{wk0}(c), for this second set of parameters and after a finite, critical time, the dynamical Landau potential $W_{k_0}(m,t)$ develops a singular point (a kink) at $m_t=0$ that persists for all (finite) later times, but vanishes asymptotically in the infinite-time limit. Through \eqnref{wk}, this kink is traced back to a singular point of $V_{k_0}(m,t)$ at $m=0$, which has the same origin as the kink in the (untilted) magnetisation rate function $V_{k=0}(m,t)$~\cite{Mei22a}, shown in \Figref{pp_magnetisation}(b). This suggests that the finite-time dynamical phase transitions of the magnetisation $m$ and of the exchanged heat $\msc{Q}$ are complementary phenomena: The transition in $\msc{Q}$ is present only when the phase transition in $m$ is absent in $V_{k_0}(m,t)$.
\subsubsection{Occurrence and continuity of transition}
We explain our previous observations by analysing the phase transition for small $m_t$. In particular, we show that the transition is continuous and determine the $\beta$-$\bq$ parameter space where it occurs. Our main strategy for this section is to assume a continuous transition of $W_{k_0}(m,t)$ at $\tcenv$ and to justify this assumption \textit{a posteriori}.
\begin{figure}
	\centering
	\includegraphics[
	width=9cm
	]{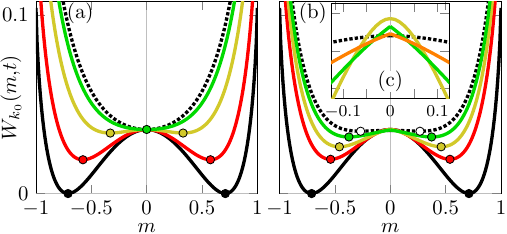}
	\caption{Dynamical Landau potential $W_{k_0}(m,t)$ at different times for varying quench parameters (coloured lines). Bullets show the minima $\pm m_t$. (a) For $\beta=5/(4J)$ and $\bq=1/(2J)$ at times $t/\tau=0$ (black), $0.25$ (red), $1$ (yellow), $1.5$ (green) and $\infty$ (dotted). (b) For $\beta=5/(4J)$ and $\bq=4/(5J)$ at times $t/\tau=0$ (black), $1$ (red), $2$ (yellow), $4$ (green) and $\infty$ (dotted). (c) Magnification of the local maximum of $W_{k_0}$ around $m=0$ in \Figref{wk0}(b), including $W_{k_0}(m,t)$ for $t/\tau=6$ (orange).}\figlab{wk0}
\end{figure}

Since $m_t=0$ for $t>\tcenv$, all continuous transitions occur at small $m$. Expanding $W_{k_0}(m,t)$ around $m=0$ gives
\algn{\eqnlab{wtaylor}
	W_{k_0}(m,t) \sim 	W_{k_0}(0,t) + \partial_m^2 W_{k_0}(0,t) \frac{m^2}2+\partial^4_m W_{k_0}(0,t) \frac{m^4}{4!}\,,
}
for $m\ll1$. A continuous dynamical phase transitions at time $\tcenv$ requires that $\partial_m^2 W_{k_0}(0,t)$ changes sign, while $\partial^4_m W_{k_0}(0,t)$ remains positive. This ensures that $W_{k_0}(0,t)$ passes from a single into a double-well.

To show that this is the case, we again recall that $W_{k_0}(m,t)$ is a function of the tilted rate function $V_{k_0}(m,t)$ through \eqnref{wk}, and that $V_{k_0}(m,t)$ obeys the Hamilton-Jacobi equation~\eqnref{hj2} with initial condition~\eqnref{vbound}. Taking partial derivatives of \eqnref{hj2} with respect to $m$, and evaluating at $m=0$, we find an exact, closed set of evolution equations for the derivatives of $V_{k_0}(m,t)$, $z_{k_0}(t) \equiv \partial_m^2V_{k_0}(0,t)$ and $w_{k_0}(t) \equiv \partial_m^4V_{k_0}(0,t)$, that we later relate to \eqnref{wtaylor}. The equations read
\sbeqs{\eqnlab{zweqs}
\algn{
	\tau\dot z_{k_0} &= 4 z_{k_0} J(\bc - \bq) - 4 z_{k_0}^2\,,\eqnlab{zeqn}\\
	\tau\dot w_{k_0} &= 4w_{k_0} [\dd{t}\log z_{k_0} -2J(\bc-\bq)]+ \dot z_{k_0}\left\{\dot z_{k_0}-2 [(\bq J-2) \bq J-2]\right\} -16 z_{k_0}\eqnlab{weqn}\,,
}
with initial conditions following from \eqnref{vbound},
\algn{\eqnlab{zwbound}
	z_{k_0}(0) = J(\bc-\bb) \,,\quad w_{k_0}(0) = 2\,,
}
}
where $\bb$ denotes the arithmetic mean $\bb\equiv (\beta+\bq )/2$ of $\beta$ and $\bq$. For later reference, we note that when
\algn{\eqnlab{qcrit}
	\bb < \bc\,,
}
then $z_{k_0}(0)>0$, and $z_{k_0}(0)\leq0$ otherwise.

The evolution equations~\eqnref{zeqn} and~\eqnref{weqn} can be solved explicitly, leading to a complicated expression for $w_{k_0}$ that we find unenlightening. However, the dynamics is easily understood qualitatively, by considering the phase portrait of the combined flow of~\eqnref{zeqn} and~\eqnref{weqn}. 
\begin{figure}
	\centering
	\includegraphics[
	width=9cm
	]{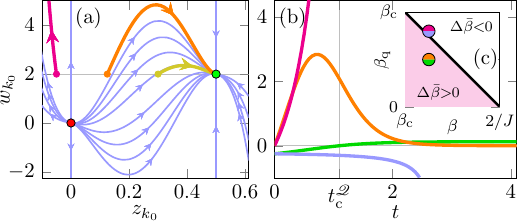}
	\caption{(a) Phase portrait of the $z_{k_0}-w_{k_0}$ dynamics~\eqnref{zeqn} and~\eqnref{weqn}, featuring the unstable (red bullet) and stable (green bullet) fixed points. Sample trajectories are shown in different colours with arrow heads indicating the direction of the flow. (b) $\partial_m^2 W_{k_0}(0,t)$ (blue and green) and $\partial_m^4 W_{k_0}(0,t)-2$ (orange and magenta) for $\bq = 1/(2J)$, and $4/(5J)$, respectively, and fixed $\beta=5/(4J)$. (c) Occurrence of the finite-time dynamical phase transition in the $\beta$-$\bq$ parameter space. Bullet colours indicate the parameter values of the $z_{k_0}$ and $w_{k_0}$ trajectories in \Figref{zwenv_portrait}(b).}\figlab{zwenv_portrait}
\end{figure}
Figure~\figref{zwenv_portrait}(a) shows the phase portrait of~\eqnref{zeqn} and~\eqnref{weqn}, featuring an unstable fixed point (red bullet) at $(z_{k_0},w_{k_0})=(0,0)$ and a stable fixed point (green bullet) at $(z_{k_0},w_{k_0})=[(\bc-\bq)J,2]$. The arrow heads indicate the time direction of the flow. We observe that all initial conditions with $z_{k_0}(0)>0$ are attracted by the stable fixed point. This is shown by the orange and yellow example trajectories. By contrast, initial conditions with $z_{k_0}(0)<0$ escape to infinity $(z_{k_0},W_{k_0})\to(-\infty,\infty)$ in finite time, exemplified by the red trajectory in \Figref{zwenv_portrait}(a).

Returning to \eqnref{wtaylor}, we express the derivatives of $W_{k_0}(m,t)$ in terms of $z_{k_0}$ and $w_{k_0}$:
\sbeqs{\eqnlab{dweqs}
\algn{
	\partial_m^2 W_{k_0}(0,t) =& z_{k_0}(t)-(\beta-\bq) J/2\,,\eqnlab{Wzrel}\\
	\partial_m^4 W_{k_0}(0,t) =& w_{k_0}(t)\,.
}
}
When $\partial_m^4 W_{k_0}(0,t)>0$ and $\partial_m^2 W_{k_0}(0,t)$ changes sign, from negative to positive, say, then $W_{k_0}(m,t)$ transitions from a double to a single well, marking a continuous finite-time dynamical phase transition. To understand for which parameters this happens, it is convenient to introduce
\algn{\eqnlab{distances}
	\Delta\bb \equiv \bc - \bb\,,\qquad \Delta\bq \equiv \bc -\bq\,.
}
For the disordering quenches ($\beta>\bc$, $\Delta\bq>0$) we consider here, $\partial_m^2W_{k_0}(0,t)$ is initially negative, $\partial_m^2W_{k_0}(0,0) = -(\beta-\bc)J<0$. This means that for any continuous transition, $\partial_m^2W_{k_0}(0,t)$ must evolve from negative to positive. When $\Delta\bb>0$, then $z_{k_0}(0)>0$ [recall \eqnref{qcrit}], so that $z_{k_0}(t)$ approaches the stable fixed point, leading to a positive $\partial_m^2W_{k_0}(0,\infty)=\Delta\bb>0$ in the long-time limit. Hence, for these parameters $\partial_m^2W_{k_0}(0,t)$ transitions from negative to positive in finite time. This is the case for $W_{k_0}(m,t)$ shown in \Figref{wk0}(a), where $\Delta\bb=1/(8J)>0$.

For $\Delta\bb\leq0$, $z_{k_0}(t)$ and $w_{k_0}(t)$ run into a finite-time divergence and no transition occurs, which is the case depicted in \Figref{wk0}(b), where $\Delta\bb=-1/(40J)<0$. The finite-time divergences of $z_{k_0}(t)$ and $w_{k_0}(t)$ reflect the formation of the kink in $W_{k_0}(m,t)$ at $m=0$, depicted in \Figref{wk0}(c).

Figure~\figref{zwenv_portrait}(b) shows the time evolution of $\partial_m^2W_{k_0}(0,t)$ and $\partial_m^4W_{k_0}(0,t)$ for disordering quenches with the parameter sets of \Figsref{wk0}(a) and \figref{wk0}(b). For the first set (green and orange lines, $\Delta\bb>0$), $\partial_m^2W_{k_0}(0,t)$ is initially negative, but changes sign at $\tcenv$. Because of the structure of the initial and final equilibrium rate functions of the model, given in \eqnref{veq} and \eqnref{freeen}, $\partial_m^4W_{k_0}(0,t)=2$ for $t=0$ and as $t\to\infty$. Nontrivially, however, $\partial_m^4W_{k_0}(0,t)\geq2$ during the entire time evolution. This ensures that, close to $\tcenv$, the finite-time dynamical phase transition is completely characterised by the expansion in \eqnref{wtaylor}, and justifies our initial assumption that the phase transition is continuous.

For the second set of parameters in \Figref{zwenv_portrait}(b) (blue and magenta, $\Delta\bb<0$), by contrast, $\partial_m^2W_{k_0}(0,t)$ remains negative, and both $\partial_m^2W_{k_0}(0,t)$ and $\partial_m^4W_{k_0}(0,t)$ diverge in finite time, when the kink in \Figref{wk0}(c) forms. In this case, the finite-time dynamical phase transition is absent.

From our small-$m_t$ analysis, we conclude that the dynamical phase transition is continuous and that it requires quenches with $\Delta\bb>0$. The coloured region in \Figref{zwenv_portrait}(c) shows where in the $\beta$-$\bq$ parameter space the finite-time dynamical phase transition occurs, i.e., where $\Delta\bb>0$. The bullets correspond to the parameter values of the plots in \Figref{zwenv_portrait}(b): While the phase transition occurs for the first set of parameters (orange and green), it is absent for the second set (blue and magenta).
\subsubsection{Critical time}\seclab{crittime}
With all necessary methods in place, we now compute the critical time $\tcenv$ for the finite-time dynamical phase transition. When $\partial_m^4W_{k_0}(0,t)>0$, as we observed, the critical time $\tcenv$ for disordering quenches is determined by the time at which $\partial_m^2W_{k_0}(0,t)$ changes sign, i.e., $\partial_m^2W_{k_0}(0,\tcenv)=z_{k_0}(t)-(\beta-\bq) J/2=0$. The solution of \eqnref{zeqn} is given explicitly by
\algn{\eqnlab{zksol}
	z_{k_0}(t)=\frac{J \Delta\bq \Delta\bb}{\Delta\bb+\Delta\bq e^{-4 J  \Delta\bq\frac{t}{\tau }}/2}\,.
}
The critical time $\tcenv$ follows from \eqnref{zksol} by setting $z_{k_0}(\tcenv)=(\beta-\bq)J/2$. Solving for $\tcenv$ gives
\algn{\eqnlab{tcenv}
	\tcenv =  \frac{\tau}{2J\Delta\bq}\log \left(\frac{\bb}{\Delta\bb}\right)\,.
}
For the parameter values in \Figsref{scgf_senv} and \figref{rf_senv}, we have $\tcenv/\tau = \ln(3)\approx1.0986$, in excellent agreement with the numerics.

Our analysis shows in particular, that $\tcenv$ is different from the critical time~\cite{Mei22a,Erm10}
\algn{\eqnlab{tcm}
	\tc = \frac{\tau}{4J\Delta\bq}\log \left(\frac{\Delta \bq}{\beta-\bc}\right)\,,
}
for the finite-time dynamical phase transition in the magnetisation $m$. Note also that $\tcenv$ in \eqnref{tcenv} diverges both when $\Delta\bq\to0$ and when $\Delta\bb\to0$, which mark the boundaries of the coloured region in \Figref{zwenv_portrait}(c)
\subsubsection{Dynamical order parameter}
We now discuss the behaviour of the dynamical order parameter in the vicinity of the transition, derive the dynamical critical exponent $=1/2$, and compare our results to direct numerical simulations of \eqnref{mastereqn}.

Close to $\tcenv$, the order parameter becomes small, $m_t\ll1$, and the expansion in \eqnref{wtaylor} is exact. Hence, we may compute $m_t$ as the minimum of \eqnref{wtaylor}. This gives
\algn{\eqnlab{dop}
	m_t \sim \css{
	\pm [-z_{k_0}(t)+J(\beta-\bq)]^{1/2}[w_{k_0}(t)]^{-1/2}\,,	&	t<\tcenv\\
	0\,,	&	t\geq\tcenv}\,,
}
i.e., a continuous, finite-time dynamical phase transition characterised by $m_t$. Close to criticality, for $|t-\tcenv|/\tau\ll1$ and $t<\tcenv$, we have $-z_{k_0}(t)+J(\beta-\bq)\propto (\tcenv-t)$. We therefore find $m_t\propto|t-\tcenv|^{1/2}$, i.e., a dynamical critical exponent of mean-field type, the same as for $m_0$ in Ref.~\cite{Mei22a}. 

As explained previously, $m_t = q^*_{k_0}(t)$ represents the most likely final magnetisation that realises $\msc{Q}=0$ in time $t$. This allows us to obtain an independent numerical estimate of $m_t$ by means of direct numerical simulations of \eqnref{mastereqn} at large but finite $N$. To this end, we generate a large number $\sim10^8$ of trajectories, and condition them on $\msc{Q}=0$ at different times $t$. We then collect the histograms of the final magnetisations $m_t$ for each $t$ and join them into one plot such that the maximum in each time slice is normalised to unity.

\begin{figure}
	\centering
	\includegraphics[
	width=9cm
	]{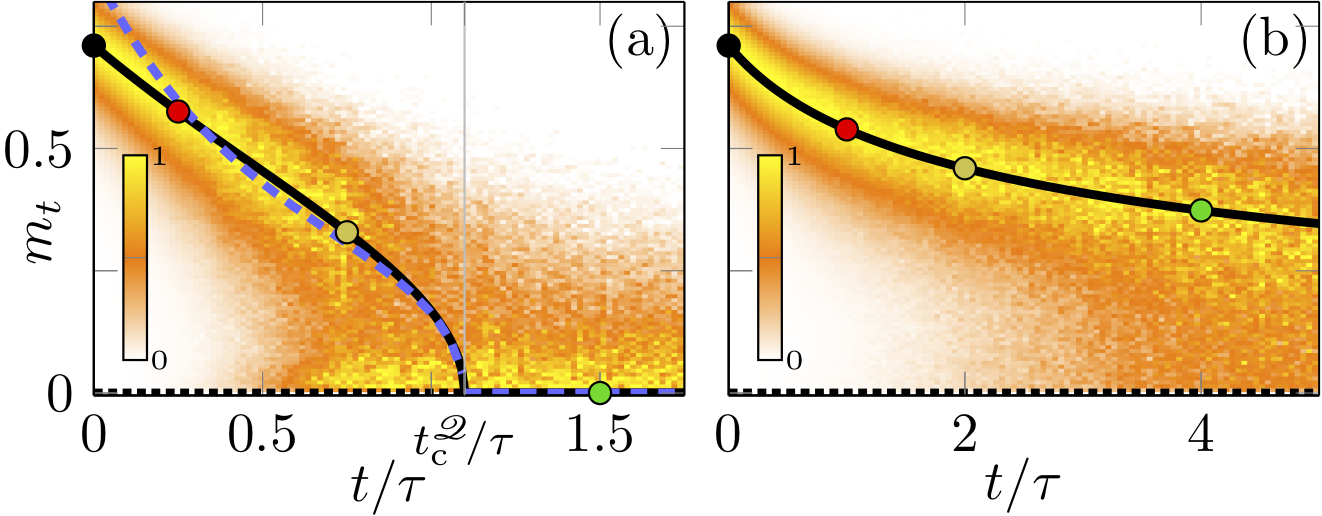}
	\caption{(a) Order parameter for $\Delta\bb>0$ with $\beta=5/(4J)$ and $\bq = 1/(2J)$. Density from numerical simulation (heat map, $N=250$, $10^8$ trajectories) and $m_t$ from theory (solid line). The dotted line shows the inactivity solution (sub-leading for $t<\tcenv$), the bullets are the same as in \Figref{wk0}(a). (b) Order parameter for $\Delta\bb<0$ with $\beta=5/(4J)$ and $\bq = 4/(5J)$. Density from numerical simulation (heat map, $N=250$, $10^8$ trajectories) and $m_t$ from theory (solid line). The dotted line shows the sub-leading inactivity solution, the bullets are the same as in \Figref{wk0}(b).}\figlab{order_parameter}
\end{figure}

Figure~\figref{order_parameter}(a) shows the so-obtained order parameter density for $\Delta\bb>0$ as a heat map. The theoretical prediction $m_t=q^*_{k_0}(t)$ is shown as the solid line. The dashed, blue line shows \eqnref{dop}, which coincides with the exact $m_t$ close to $\tcenv$, but deviates for short times. The bullets correspond to the minima of the Landau potentials $W_{k_0}(m,t)$ in \Figref{wk0}(a). We observe good agreement between the yellow regions of high order parameter density with $m_t$ calculated from the optimal fluctuations. The transition between finite $m_t$ at $t\leq\tcenv$ and $m_t=0$ for $t>\tcenv$ is clearly visible. Note that the transition is inverted (finite to zero) compared to the transitions of $\bar m$ and $m_0$ shown in \Figsref{phase_diag}(b) and \figref{pp_magnetisation}(c) (zero to finite). Close to $\tcenv$, we observe strong fluctuations of $m_t$ and a high order-parameter density at $m_t\approx 0$ (black dotted line) even for $t\approx0.7\tau<\tcenv$. This is a finite-$N$ effect, as we explain in more detail in the next section.

Figure~\figref{order_parameter}(b) shows the same as \Figref{order_parameter}(a) but now for a parameter set with $\Delta\bb<0$. Here, the bullets correspond to the minima of $W_{k_0}$ in \Figref{wk0}(b). We observe no phase transition, as the order parameter remains finite at all times and approaches zero asymptotically. 

In both \Figref{order_parameter}(a) and \Figref{order_parameter}(b), the numerical data turns noisier for increasing $t$, because the event $\msc{Q}=0$ becomes less typical as the spin system equilibrates with the environment. Consequently, less trajectories remain after conditioning on $\msc{Q}=0$, resulting in an increased statistical error.

\subsubsection{Optimal fluctuations}\seclab{optfluct}
Finally, the origin of the dynamical phase transition can be viewed from the perspective of the optimal fluctuations that generate $\msc{Q}=0$. To improve our intuition for these fluctuations, it is useful to consider how the condition $\msc{Q}=0$ constrains their dynamics.

According to \eqnref{legda}, the optimal fluctuation $q^*_{k_0}(s)_{0\leq s\leq t}$ for $\msc{Q}=0$ must satisfy
\algn{\eqnlab{consten}
	\msc{E}[q^*_{k_0}(t)]=\msc{E}[q^*_{k_0}(0)]\,,
}
i.e., the internal energy before the quench and at time $t$ must coincide. For $\msc{E}(m) = -\bq J m^2/2$, \eqnref{consten} translates into
\algn{\eqnlab{trajconstr}
	q^*_{k_0}(t) = \pm q^*_{k_0}(0)\,.
}
Hence, the requirement that $\msc{Q}=0$ forces the initial and end points of the optimal fluctuations to agree up to a sign.

The constraint \eqnref{trajconstr} on the optimal fluctuation gives a simple qualitative explanation of why the dynamical phase transition occurs at a finite time, by considering the most likely ways to achieve $\msc{Q}=0$ at short and long times: For short times $t\ll\tau$, the most likely way  is to start and end close to the most likely initial condition $q^*_{k_0}(0)\approx\bar m(\beta)>0$, because the relaxation dynamics can be sustained for short times at low probabilistic cost.  For long times $t\gg\tau$ and $\Delta\bb>0$, by contrast, the system is more likely to start at vanishing magnetisation, at high initial probabilistic cost, because it is also the most likely \textit{final} magnetisation, i.e., $\bar m(\bq) =0$. In other words, although the initial probabilistic cost of $q^*_{k_0}(0)\approx0$ is high, the system may then stay close to the origin for an arbitrary amount of time at no additional cost.

According to this argument one expects different optimal fluctuations for short and long times, implying a transition between the two behaviours at some intermediate time, given by the critical time $\tcenv$.

For $\Delta\bb<0$, the probability of initiating (and staying) at $m=0$ is always too low, compared to starting (and ending) somewhere in the middle ground between to a likely initial condition and an unlikely final condition.

At finite $N$, the variable sign in \eqnref{trajconstr} gives rise to the premature transition observed in the numerics in \Figref{order_parameter}(a). This is, because at any finite $N$, trajectories that initiate close to $m=0$ have two possibilities that occur with similar probability: At time $t$ they may end up at either the positive or negative value of their initial magnetisation. Trajectories that initiate far away from the origin, close to the initial minima of $\veq(m)$, say, have effectively only one possibility, $q^*_{k_0}(t) = q^*_{k_0}(0)$, because the probability of trajectories crossing the origin and ending up at their negative initial magnetisation, $q^*_{k_0}(t) = -q^*_{k_0}(0)$, is exponentially suppressed. As a result, the probability to start and end close to the origin is enhanced at finite $N$.

This effect becomes smaller as $N$ increases, since trajectories for which both possibilities in \eqnref{trajconstr} are of similar probability, recide closer and closer to the origin. In the thermodynamic limit, only the $+$ constraint in \eqnref{trajconstr} survives, so that the optimal fluctuations always obey $q^*_{k_0}(t) = q^*_{k_0}(0)$, see~\eqnref{timerev}. Based on this argument, we have checked that the premature transition in \Figref{order_parameter}(a) is absent when we enforce $q^*_{k_0}(t) = q^*_{k_0}(0)$ also at finite $N$.

From our direct numerical simulations for $\Delta\bb>0$, we visualise the optimal fluctuations by conditioning the trajectories on $\msc{Q}=0$ at times smaller and  larger than $\tcenv$. In order to emulate the thermodynamic limit, we now enforce $q^*_{k_0}(t) = q^*_{k_0}(0)$, instead of admitting both signs in \eqnref{trajconstr}. Tracking the entire history of the conditioned trajectories provides us with a numerical estimate of the conditioned trajectory density in the thermodynamic limit, before and after the phase transition. The resulting trajectory densities, normalised to unity for each time slice, are shown in \Figref{heat_trajectories}.
\begin{figure}
	\centering
	\includegraphics[
	width=9cm
	]{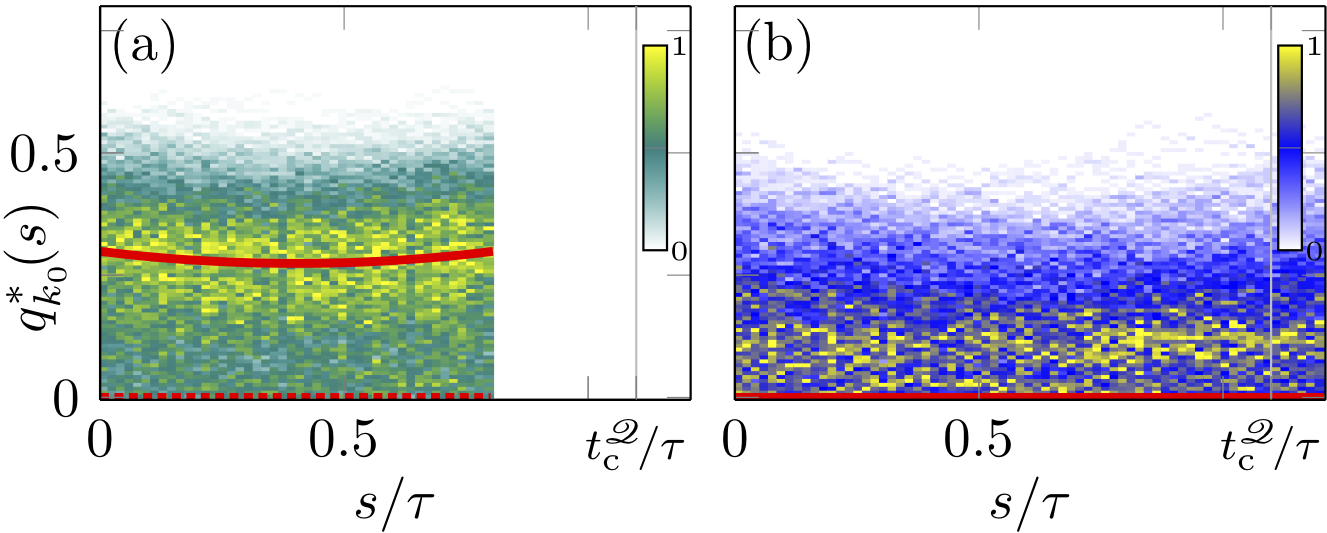}
	\caption{Optimal trajectories for $\msc{Q}(t)=0$ with $\beta=5/(4J)$ and $\bq=1/(2J)$, constrained by \eqnref{trajconstr} with only the $+$ sign. (a) Trajectory density from simulation (heat map, $N=250$, $10^8$ trajectories) and optimal fluctuation (red line) from theory for  $t=0.8\tau<\tcenv$. The dashed line shows the sub-leading fluctuation. (b) Same as in \Figref{heat_trajectories}(a) but with $t=1.2\tau>\tcenv$.}\figlab{heat_trajectories}
\end{figure}

Figure~\figref{heat_trajectories}(a) shows the trajectory density (heat map) for $t < \tcenv$, together with the corresponding optimal fluctuation $q^*_{k_0}(s)_{0\leq s \leq t}$ (red line). The dotted line shows the sub-leading fluctuation $q(s)=0$. Although the numerical data is noisy for the reasons mentioned previously, we observe good agreement between the yellow streak of high trajectory density and the theoretical curve. Note that there is a negative, but otherwise identical optimal fluctuation, not shown in \Figref{heat_trajectories}, that starts and ends at a finite negative magnetisation. For $t > \tcenv$ shown in \Figref{heat_trajectories}(b), by contrast, the optimal fluctuation remains zero at all times, reflected in both the numerics and the theory. 
\section{Conclusions}\seclab{conc}
Combining elements from stochastic thermodynamics and large-deviation theory, we derived a powerful extension of the Hamiltonian method for computing the time-dependent statistics of thermodynamic observables after an instantaneous temperature quench. The approach proves particularly effective for the analysis of finite-time dynamical phase transitions, as it naturally gives rise to a dynamical generalisation of Landau theory. The corresponding dynamical Landau potential allows for an unambiguous identification of the dynamical order parameter and of the associated dynamical phases in the phase diagram. Our theory applies to systems with underlying stochastic dynamics that admit well-defined thermodynamic or weak-noise limits.

We introduced our approach using the Curie-Weiss spin model as a concrete, non-trivial example of a system with an equilibrium phase transition. For disordering quenches across the phase boundary, the magnetisation $m$ of this system was shown to exhibit a finite-dynamical phase transition in Ref.~\cite{Mei22a}. Using our new method, we conducted a detailed analysis of the statistics of the heat $\msc{Q}$ released into the environment after such a disordering quench.

In a finite region of the parameter space, our investigation revealed another finite-time dynamical phase transition associated with this observable. The transition manifests itself in a finite-time kink in the probability distribution of $\msc{Q}$ and classifies as continuous, with mean-field critical exponent, similar to the transition for $m$~\cite{Mei22a}. Apart from these similarities, however, the two transitions exhibit very different properties.

On the trajectory level, we showed that the new finite-time dynamical phase transition associated with $\msc{Q}$ is related to a constraint on the initial and end points of individual trajectories. The most likely ways to satisfy the constraint differ in the short- and long-time limits. This implies the occurrence of a sudden switch in the optimal, most likely fluctuation at finite time and thus provides a qualitative explanation for the occurrence of the phase transition. At finite $N$, we argued that the constraint posed on the fluctuations is effectively weaker for trajectories that reside close to $m=0$, which explains a premature phase transition at $t<\tcenv$ observed in our direct numerical simulations.

The extended Hamiltonian method presented here opens the door to a complete, finite-time analysis of the stochastic thermodynamics of systems subject to quenches of either the temperature or other external parameters. Our analysis of the finite-time statistics of the released heat $\msc{Q}$ reveals that multiple finite-time dynamical phase transitions associated with different observables and generated in different ways occur in the relaxation dynamics of the same system, as a consequence of different constraints posed on the dynamics at the trajectory level. This indicates that finite-time dynamical phase transitions with distinct properties are an integral part of far-from-equilibrium relaxation processes, that occur in a wide range of physical situations. The dynamical Landau theory we propose for the study of these transitions has proven powerful in identifying the distinct time-dependent phases and for classifying them in terms of well-known equilibrium categories. We are confident that our methods will be useful in the study of finite-time dynamical phase transitions in other models and for a variety of observables.

As for the Curie-Weiss model, the next logical step is to investigate the finite-time statistics of entropy production in response to quenches. This would give a more detailed account of the finite-time dynamics of dissipation and provide further insights into the irreversibility of relaxation processes in the thermodynamic limit. The analysis is slightly more involved in this case, because the observable depends explicitly on time, leading to time-dependent constraints on the trajectories. Notwithstanding, the theory developed here applies without further limitations.

An important generalisation our method is the inclusion of steady and time-dependent driving. This enables the study of dynamical observables not only in non-equilibrium steady states but also the transient relaxation towards them. For example, the characteristic kink in the rate function of entropy production found at steady state in Refs.~\cite{Meh08,Lac08,Pro19} could have formed in the transient, as a consequence of a finite-time dynamical phase transition. An analysis of this and related problems with our methods would provide new insights into how known dynamical phase transitions are generated.

Finally, the fact that our theory applies in the thermodynamic limit, raises the question how finite $N$ as well as critical fluctuations (in both space and time~\cite{Mei22a}) affect finite-time dynamical phase transitions. In equilibrium, finite-$N$ corrections are known to potentially alter the location of the critical point and even change the order of phase transitions~\cite{Gol92}. Our numerical simulations in \Figref{order_parameter}(a) indicate that such corrections could also occur for finite-time dynamical phase transitions. How precisely these finite-$N$ corrections and critical fluctuations, responsible for corrections to mean-field critical exponents at equilibrium~\cite{Gol92,Cha95}, affect finite-time dynamical phase transitions, remains an intriguing open question.
\begin{ack}
	This work was supported by the European Research Council, project NanoThermo (ERC-2015-CoG Agreement No. 681456) and by a Feodor-Lynen Fellowship (JM) of the Alexander von Humboldt-Foundation.
\end{ack}
\appendix
\section{Fluctuation relation for \texorpdfstring{$\msc{Q}$}{Q}}\seclab{detfluctrel}
We prove a fluctuation relation for $\msc{Q}$, \eqnref{detfluctrel} in the main text. The proof given here is based on the following observation: In an equilibrium state at inverse temperature $\bq$, the joint probability $P^\text{eq}_\text{q}(m,t;m',0)$ obeys detailed balance~\cite{Kam07},
\algn{\eqnlab{eqjoint}
	P^\text{eq}_\text{q}(m,t;m',0) = P^\text{eq}_\text{q}(m',t;m,0)\,.
}
The subscript $\text{q}$ indicates that both $m$ and $m'$ are sampled from the equilibrium distribution at inverse temperature $\bq$. Conditioning on the initial equilibrium state, we now write
\sbeqs{
\algn{
	P^\text{eq}_\text{q}(m,t;m',0) =& P_\text{q}(m,t|m',0)P_\text{q}^\text{eq}(m')\,,\\
	P^\text{eq}_\text{q}(m',t;m,0) =& P_\text{q}(m',t|m,0)P_\text{q}^\text{eq}(m)\,.
}
}
Using this and \eqnref{eqjoint}, we relate the conditional probability distributions by
\algn{
	P_\text{q}(m',t|m,0) =& P_\text{q}(m,t|m',0)\frac{P_\text{q}^\text{eq}(m')}{P_\text{q}^\text{eq}(m)}\,,\eqnlab{eqrelation0}\\
			 =& P_\text{q}(m,t|m',0)\ee^{-N\bq[\msc{F}(\bq,m')-\msc{F}(\bq,m)]}\,.\eqnlab{eqrelation}
}
Swapping the summation indices $m\leftrightarrow m'$ in \eqnref{mgf1} we obtain for $G(k,t)$
\algn{
	G(k,t) 	=& \sum_{m,m'}\ee^{-Nk\bq\left[\msc{E}(m')-\msc{E}(m)\right]} P_\text{q}(m',t|m,0)P^\text{eq}(m,0)\,.
}
Using \eqnref{eqrelation} and rearranging the terms gives
\algn{
	G(k,t)	= \sum_{m,m'}\ee^{-N\bq(-k+\beta/\bq-1)[\msc{E}(m)-\msc{E}(m')]}P_\text{q}(m,t|m',0)P^\text{eq}(m',0)\,.
}
Comparing with \eqnref{mgf1}, and shifting $k$ by $k_0 =  (\beta/\bq-1)/2$, gives the finite-time fluctuation relation
\algn{\eqnlab{fluctmgfa}
	G(k+k_0,t) = G(-k + k_0,t)\,,
}
which shows that $G(k,t)$ is symmetric about $k=k_0$. For the probability distribution $P(\msc{Q},t)$, \eqnref{fluctmgfa} implies
\algn{\eqnlab{fluctdista}
	P(\msc{Q},t) = P(-\msc{Q},t)\ee^{-N(\beta/\bq-1)\msc{Q}}\,,
}
leading to
\algn{
	I(\msc{Q},t)-I(-\msc{Q},t) = (\beta/\bq-1)\msc{Q}\,,
}
for the rate function.
\section{Variational principle for \texorpdfstring{$\Lambda(k,t)$}{L(k,t)}}\seclab{variation}
Here we prove a variational principle for $\Lambda(k,t)$ valid for all $k$ and $t$. We consider the variation  $\delta\Lambda(k,t)$ with respect to $q^*_k(s)$ and $p^*_k(s)$. From the integral expression \eqnref{lamk} we obtain
\begin{multline}
	\delta \Lambda(k,t) = -\int_0^t \ed s \big[ \delta p^*_k \dot q^*_k + p^*_k \dd{s} \delta q^*_k  - \partial_q\msc{H}(q^*_k,p^*_k)\delta q^*_k - \partial_p\msc{H}(q^*_k,p^*_k)\delta p^*_k \big]\\
	 - \delta q_k^*(0)\dd{q}\veq[q^*_k(0)] +k\{\delta q_k^*(t)\partial_q\msc{A}[q^*_k(t),t]-\delta q_k^*(0)\partial_q\msc{A}[q^*_k(0),0]\}\,.
\end{multline}
An integration by parts gives
\begin{multline}\eqnlab{variation}
	\delta \Lambda(k,t) = \int_0^t \ed s   \big[\dot p^*_k + \partial_q\msc{H}(q^*_k,p^*_k) \big]\delta q^*_k(s)-\int_0^t \ed s   \big[\dot q^*_k- \partial_p\msc{H}(q^*_k,p^*_k)\big]\delta p^*_k(s)\\
	 +\delta q^*_k(t)\big\{- p^*_k(t) + k\partial_q\msc{A}[q^*_k(t),t]\big\}+\delta q^*_k(0)\big\{p^*_k(0)- k\partial_q\msc{A}[q^*_k(0),0] - \dd{q}\veq[q^*_k(0)]\big\}\,.
\end{multline}
Applying the Hamilton equations \eqnref{heom} together with the boundary conditions \eqnref{psbound} we readily obtain $\delta \Lambda(k,t) = 0$. Equation~\eqnref{variation} can be conveniently written as
\algn{
	\delta \Lambda(k,t) = \int_0^t \ed s   \left[\frac{\delta \Lambda}{\delta p^*_k} \delta p^*_k +  \frac{\delta \Lambda}{\delta q^*_k} \delta q^*_k\right]\,,
}
where the variational derivatives
\sbeqs{\eqnlab{varder}
\algn{
	\frac{\delta \Lambda}{\delta p^*_k} =& -\dot q^*_k+ \partial_p\msc{H}(q^*_k,p^*_k)\,,\\
	\frac{\delta \Lambda}{\delta q^*_k}  =& \dot p^*_k+ \partial_q\msc{H}(q^*_k,p^*_k)+\delta(s-t)\left[- p^*_k + k\partial_q\msc{A}(q^*_k,s)\right]\\
	&+\delta(s)\left[p^*_k- k\partial_q\msc{A}(q^*_k,s) - \dd{q}\veq(q^*_k)\right]\nn\,,
}
}
vanish individually, i.e., $\frac{\delta \Lambda}{\delta p^*_k} = \frac{\delta \Lambda}{\delta q^*_k} = 0$, as stated in the main text.
\end{document}